\documentclass{article}

\usepackage{arxiv}

\usepackage[utf8]{inputenc} 
\usepackage[T1]{fontenc}    
\usepackage{hyperref}       
\usepackage{url}            
\usepackage{booktabs}       
\usepackage{amsfonts}       
\usepackage{nicefrac}       
\usepackage{microtype}      
\usepackage{lipsum}		
\usepackage{graphicx}
\usepackage{natbib}
\usepackage{doi}
\usepackage{amsmath}

\title{Modeling Brain MRI Using Persistent Homology and Multilevel Functional Data Analysis}

\author{{\includegraphics[scale=0.06]{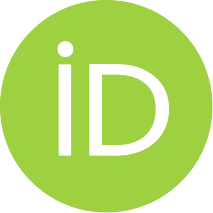}\hspace{1mm}Shashipraba N. K. Rajakaruna}\\
	Department of Mathematics and Statistics,\\
	Texas Tech University\\
	Texas, USA \\
	\And
	\href{https://orcid.org/0000-0001-7037-2721}{\includegraphics[scale=0.06]{orcid.pdf}\hspace{1mm}Asim K. Dey} \\
	Department of Mathematics and Statistics,\\
	Texas Tech University\\
	Texas, USA \\
    \And
    	{\includegraphics[scale=0.06]{orcid.pdf}\hspace{1mm}A. Alexandre Trindade} \\
	Department of Mathematics and Statistics,\\
	Texas Tech University\\
	Texas, USA \\
}

\renewcommand{\shorttitle}{\textit{arXiv} Template}

\hypersetup{
pdftitle={A template for the arxiv style},
pdfsubject={q-bio.NC, q-bio.QM},
pdfauthor={David S.~Hippocampus, Elias D.~Striatum},
pdfkeywords={First keyword, Second keyword, More},
}

\begin{document}
\maketitle

\begin{abstract}
	Persistent homology provides a multiscale representation of biomedical images by capturing higher-order topological features that reflect their underlying structural organization. However, the resulting topological summaries are typically used as predictors or features for classification and group comparisons rather than treated as primary variables of interest. We construct a generalized multilevel functional framework for analyzing persistent-homology summaries as longitudinal functional responses in repeated three-dimensional structural magnetic resonance imaging (MRI). Specifically, we represent topological features using Betti curves and model these curves as count-valued functional responses. A negative-binomial distribution accommodates the discrete and potentially overdispersed nature of Betti counts, while the multilevel formulation accounts for the dependence induced by repeated measurements and separates between-subject and within-subject sources of functional variation. Functional principal component analysis further evaluates the dominant modes of variation at each level. A Bayesian approach is used for joint estimation of the functional regression and multilevel functional principal components. We apply this modeling framework to longitudinal structural MRI data from the OASIS-2 study to investigate associations between brain topology and demographic and clinical characteristics, including age, gender, follow-up time, and dementia severity. The results demonstrate that the framework can capture covariate-associated variation across the filtration continuum while evaluating distinct sources and patterns of longitudinal variation across homology dimensions. 
\end{abstract}


\keywords{persistent homology \and Betti curves \and functional data analysis \and
multilevel functional data \and brain MRI \and Alzheimer's disease}

\section{Introduction}\label{intro}

Neurodegenerative brain disorders, such as Alzheimer's disease, are accompanied by progressive structural changes in the brain that evolve across the course of the disease. 
Structural magnetic resonance imaging (MRI) provides a noninvasive means of characterizing these changes and plays an important role in identifying patterns of neurodegeneration across the disease continuum~\cite{frisoni2010clinical, jack2018nia, knopman2021alzheimer}. Measures derived from structural MRI, including regional brain volumes, hippocampal atrophy, and cortical thickness, provide important quantitative information for assessing cerebral atrophy and its progression. Beyond these established morphometric measures, the spatial organization and shape of brain structures provide complementary information about disease-related morphological changes associated with neurodegeneration~\cite{singh2023topological}.\\

Topological data analysis (TDA) provides a rigorous mathematical framework for quantifying the shape and structural organization of complex data through topological features that persist across multiple scales~\cite{Carlsson2009Topology, Ghrist2008Barcodes}. This framework offers a complementary perspective to conventional morphometric measures by capturing aspects of spatial organization and morphological complexity that may not be fully reflected by measures such as regional volume or cortical thickness. Persistent homology, a central method in TDA, tracks the appearance and disappearance of topological features across a filtration~\cite{Edelsbrunner2002Persistence,chazal2021introduction}. A filtration consists of a nested sequence of topological spaces or complexes indexed by a threshold or scale parameter and provides a multiscale representation of the structural organization of the data. By retaining topological information over a full range of filtration values, persistent homology provides a topological signature of how structural organization changes with scale, rather than representing the data at a single threshold. This multiscale perspective is particularly relevant for brain MRI, where spatial arrangement, connectivity, and morphological complexity may provide additional information for understanding disease-related changes in brain structure~\cite{singh2023topological}. Several filtration methods and topological complexes have been developed to extract multiscale topological descriptors from complex data. For images defined on regular pixel or voxel grids, the cubical complex~\cite{kaczynski2006computational} provides a natural representation that preserves the spatial and topological structure of the underlying image. The cubical complex enables persistent homology to be computed directly from image data, allowing topological features to be examined across multiple scales.\\

Recent biomedical applications have demonstrated that persistent homology based topological descriptors can effectively capture complex image morphology and can be used to investigate associations with clinically relevant outcomes~\cite{Moon2023Persistent, singhal2026topology, shiraishi2026persistent}. In brain network analysis, recent studies highlight the utility of topological measures for evaluating structural brain organization, revealing complex connectivity patterns and nonlinear changes in network topology across the human lifespan~\cite{Songdechakraiwut2023, Chung2024, Mousley2025Topological}. In Alzheimer's disease, persistent homology based summaries have identified changes in both structural and functional brain organization, including altered white-matter connectivity and disrupted functional-network integration across disease stages~\cite{Kuang2020WhiteMatter, Xing2022Spatiotemporal}. In addition, \citet{SaadatYazdi2021BettiCurves} use persistent homology to examine age- and Alzheimer’s-related changes in brain morphology, showing that the resulting topological summaries contain information useful for distinguishing disease-related structural patterns from those associated with normal aging. Related studies further underscore the utility of topological representations for investigating Alzheimer’s-related alterations in brain organization and disease progression through functional-network topology and structural-connectome models~\cite{Xu2024TopologyClustering, Goodbrake2024BrainChains, yi2025topological}.\\

Despite the growing use of persistent homology in biomedical data analysis, particularly in brain imaging, the resulting topological summaries are often used as features for prediction, classification, or group comparison rather than treated as primary variables of interest in statistical modeling and inference. In many existing applications, persistent homology representations are reduced to scalar or vector-valued summaries and subsequently incorporated into conventional statistical or machine-learning frameworks, with the primary objective of distinguishing disease groups, predicting clinical outcomes, or identifying topological differences across populations. Although these approaches demonstrate the potential of topological descriptors for capturing structural characteristics that may be overlooked by conventional imaging measures, they do not fully exploit the functional nature of topological summaries produced across a continuum of filtration values. A natural approach to representing the functional topological summaries is through Betti curves. For a given homology dimension, a Betti curve quantifies the number of topological features present at each filtration value, providing a functional summary across the filtration domain rather than a single scalar measure~\cite{chazal2021introduction}. 
When brain MRI data are collected repeatedly from the same individuals, the resulting Betti curves vary across visits, yielding longitudinal functional responses that capture both within-subject changes over time and between-subject heterogeneity in brain morphology~\cite{crainiceanu2024functional}. Modeling these Betti curves provides a means to quantify changes in topological characteristics over time while accounting for the within-subject dependence induced by repeated measurements.\\

Functional data analysis provides a natural framework for modeling these longitudinal functional responses, i.e., Betti curves, while preserving variation in both the magnitude and shape of the curves. Multilevel functional principal component analysis decomposes functional variation into between-subject and within-subject components through level-specific covariance structure~\cite{di2009multilevel,10.1214/10-EJS575, shou2015structured, Lin2024Multilevel}. Functional mixed models further provide a flexible framework for relating scalar covariates to longitudinal functional responses while accounting for the dependence induced by repeated measurements~\cite{scheipl2015functional, li2022fixed}. Extensions to generalized functional responses accommodate non-Gaussian outcomes while retaining flexible covariate effects and multilevel functional random effects~\cite{10.1214/16-EJS1145, goldsmith2015generalized}. \\

In this study, we develop a multilevel functional framework for investigating changes in brain topology derived from three-dimensional structural MRI using persistent homology. 
In particular, we model the resulting Betti curves as longitudinal functional responses and evaluate how brain topology evolves across repeated visits. This formulation allows associations with baseline age, gender, follow-up time, and dementia severity to vary over the filtration continuum, thereby identifying where topological differences are more pronounced and how different regions of the Betti curves vary across these covariates. To accommodate the discrete nature and overdispersion of Betti counts, we formulate the analysis using a generalized multilevel functional model with a negative-binomial distribution, extending the generalized framework of~\citet{goldsmith2015generalized} to count-valued Betti-curve responses. The multilevel representation further separates between-subject differences in brain topology from within-subject differences and characterizes the dominant modes of variation at each level through functional principal components. In addition, we use Bayesian estimation implemented in Stan~\cite{Jiang2025BayesianFunctional}, which facilitates the joint estimation of the functional regression and principal component structures. \\

The remainder of the paper is organized as follows. Section~\ref{sec:2} describes the topological construction of the Betti curves. Section~\ref{sec:3} introduces the generalized multilevel functional regression model and its multilevel functional principal component representation, and Section~\ref{sec:4} describes the estimation procedure. Section~\ref{sec:5} presents the application to Alzheimer's disease progression and the corresponding results, followed by a discussion in Section~\ref{sec:6}.

\section{Topological representation of brain MRI}\label{sec:2}

\subsection{Persistent homology and topological summary of brain MRI}

In recent years, the growing complexity and dimensionality of neuroimaging data have challenged conventional analytical approaches, including statistical modeling, signal processing, and graph-theoretical methods. Topological data analysis, an emerging framework at the intersection of algebraic topology, computational geometry, and statistical inference, provides a complementary framework for evaluating higher-order geometric and structural organization of brain images, such as brain MRI. A central tool for evaluating such higher-order topological summaries is \emph{persistent homology}, which can be used to quantify the evolution of topological structures in brain images across multiple scales. \\

The construction of persistent homology often begins by representing the data using a \emph{simplicial complex}, denoted by $\mathcal{K}$, which provides a combinatorial description of the geometric structure of the data. A simplicial complex is constructed from basic geometric objects called \emph{simplices} of different dimensions, including vertices, edges, filled triangles, and solid tetrahedra, corresponding to dimensions $0$, $1$, $2$, and $3$, respectively. More generally, a simplicial complex is a collection of simplices satisfying the property that whenever a simplex $\sigma \in \mathcal{K}$ and $\tau \subseteq \sigma$ is a face of $\sigma$, then $\tau \in \mathcal{K}$, and for any two simplices $\sigma,\sigma' \in \mathcal{K}$, the intersection $\sigma \cap \sigma'$ is either empty or a common face of both $\sigma$ and $\sigma'$. In this way, a simplicial complex provides a mathematically coherent representation of the geometric and topological structure present in the data~\citep{Wasserman:2018,chazal2021introduction}. 
That is, persistent homology studies how the topology of a simplicial complex $\mathcal{K}$ evolves as a scale parameter $s$ varies. This evolution is represented by a \textit{filtration}, defined as a nested sequence of simplicial complexes, $\mathcal{K}_{s_0} \subseteq \mathcal{K}_{s_1} \subseteq \cdots \subseteq \mathcal{K}_{s_L},$ where $s_1 < s_2 < \cdots < s_L$. \\

For point-cloud data, one widely used filtration is the Vietoris–Rips filtration, in which simplices are added according to pairwise distances. More generally, a filtration may be induced by a function \(f:\mathcal{K} \to \mathbb{R}\) satisfying \(f(\sigma) \leq f(\tau)\) whenever \(\sigma \subseteq \tau\). At each stage of the filtration, the topological structure of the simplicial complex $\mathcal{K}_s$ is described by its {homology groups} \( H_p(\mathcal{K}_s) \). Informally, $H_p(\mathcal{K}_s)$ captures the $p$-dimensional \emph{holes} in the complex $\mathcal{K}_s$: $H_0$ represents connected components, $H_1$ represents loops or one-dimensional holes, $H_2$ represents voids or two-dimensional holes, and higher-dimensional homology groups capture their corresponding higher-dimensional analogues.  
As the filtration progresses, features such as connected components and loops may appear and disappear. The inclusion \( \mathcal{K}_{s_i} \hookrightarrow \mathcal{K}_{s_{i+1}} \) induces linear map between homology groups: $H_p(\mathcal{K}_{s_i}) \to H_p(\mathcal{K}_{s_{i+1}})$, allowing us to track the topological features across scales. A homological feature is said to be born at scale $s_{i}$ if $\mathcal{K}_{s_{i}}$ is the first complex in the filtration in which the feature appears, and dies at scale $s_{j}$ if the feature disappears or merges into an older feature at $\mathcal{K}_{s_{j}}$. Thus, each homological feature can be represented by an interval $(s_{i},s_{j})$ that records the range of scales over which the feature persists. A standard way to represent the birth-death pairs (i.e., persistence) of all homology features is the \textit{persistence diagram} ($\mathcal{D}$), which is defined as the multi-set
\begin{equation}\label{Eq:PD}
\mathcal{D}=\{(s_i, s_j) \in \mathbb{R}^{2} | s_i<s_j\}.
\end{equation}

In general, the points of a persistence diagram $\mathcal{D}$ lie above the diagonal $\mathcal{L} = \{(t, t) \mid t \in \mathbb{R}\}$ in the plane. The persistence of a point in $\mathcal{D}$ is its vertical distance from $\mathcal{L}$. Points closer to $\mathcal{L}$ have shorter lifetimes and lower persistence, while a point on the diagonal $(t, t) \in \mathcal{L}$ has zero persistence. Points farther from $\mathcal{L}$ indicate features with longer lifetimes. Longer-lived features are generally regarded as more prominent, whereas shorter-lived features are often viewed as noise~\citep{edelsbrunner2010computational,hickok2022analysis}. \\


While simplicial complexes (e.g., the Vietoris–Rips complex) are natural for unstructured data such as point clouds, cubical complexes are more appropriate for structured data defined on regular grids, including images and volumetric scans~\citep{Otter_et_al2017}. A cubical complex extends the idea of a simplicial complex by replacing simplices with unit \emph{cubes} of various dimensions: 0-cubes (points), 1-cubes (edges), 2-cubes (squares), 3-cubes (cubes), and so on. Formally, a cubical complex $\mathcal{K}$ is a finite collection of unit cubes in \( \mathbb{R}^n \) such that: (i) if a cube \( Q \in \mathcal{K} \), then all its faces (lower-dimensional cubes) are also in \( \mathcal{K} \), and (ii) the intersection of any two cubes in \( \mathcal{K}\) is either empty or a common face. This representation is particularly suitable for imaging data, because it directly preserves the underlying grid structure of pixels or voxels.  To study image topology, e.g., homology groups, an image can be represented as a function \( I : \Omega \to \mathbb{R} \), where \( \Omega \) is the image domain and \( I \) assigns an intensity value to each pixel or voxel. A filtration is then constructed by thresholding these intensities. Among the various filtrations available for images, the two standard choices are \textit{sublevel} and \textit{superlevel} set filtrations, which build nested sequences of spaces based on image intensity thresholds. In a sublevel set filtration, the cubical complex at threshold \( s \) is defined as
$\mathcal{K}_s = \{\, Q \in \Omega : I(Q) \le s \,\}$. This construction ensures that all cubes with intensity values less than or equal to $s$ are included in the complex, along with their lower-dimensional faces. As the threshold $s$ increases, cubes are progressively added, producing the nested sequence
$\mathcal{K}_{s_1} \subseteq \mathcal{K}_{s_2} \subseteq \cdots \subseteq \mathcal{K}_{s_L}$, for $s_1 < s_2 < \cdots < s_L$. In a superlevel-set filtration, cubes are instead included according to $\mathcal{K}_s = \{\, Q \in \Omega: I(Q) \ge s \,\}$. In this case, decreasing the threshold adds cubes from high-intensity to low-intensity regions, resulting in a nested sequence when $s_1 > s_2 > \cdots > s_L$. Although we use a sublevel-set filtration in this study, both constructions allow topological features to be tracked across different image intensity levels and provide a computationally efficient framework for topological analysis of imaging data~\citep{kaczynski2006computational,franccois2024train}. \\

While the persistence diagram $\mathcal{D}$ defined in Eq.~\ref{Eq:PD} summarizes topological features through their birth and death times, Betti curves provide a complementary functional representation of these features across the filtration. For a cubical filtration $\{\mathcal{K}_s\}_{s\in\mathbb{R}}$, the $p$th Betti number at a fixed threshold $s$ is defined as the rank of the $p$-th homology group, $h_p(s)=\operatorname{rank}\bigl(H_p(\mathcal{K}_s)\bigr)$ which counts the number of $p$-dimensional topological features present in the complex $\mathcal{K}_s$~\citep{edelsbrunner2010computational}. Thus, $h_0(s)$ counts the number of connected components, $h_1(s)$ counts the number of loops, and $h_2(s)$ counts the number of voids present in the complex at threshold $s$. 
Evaluating these Betti numbers over a finite grid of filtration values $s_1,\ldots,s_L$ yields the discretized $p$th Betti curve, $h_p=\bigl(h_p(s_1),h_p(s_2),\ldots,h_p(s_L)\bigr)$, which describes how the number of $p$-dimensional topological features evolves across the filtration~\cite{chazal2021introduction,Su2026Topological}. Therefore, a Betti curve provides a functional summary of how topological features appear, persist, and disappear over the filtration domain. In contrast to a persistence diagram, which is a multiset of birth--death pairs, the Betti curve represents the number of topological features as a one-dimensional function over the filtration parameter $s$. \\

\begin{figure*}[!ht]
    \centering
        \includegraphics[width=0.65\textwidth]{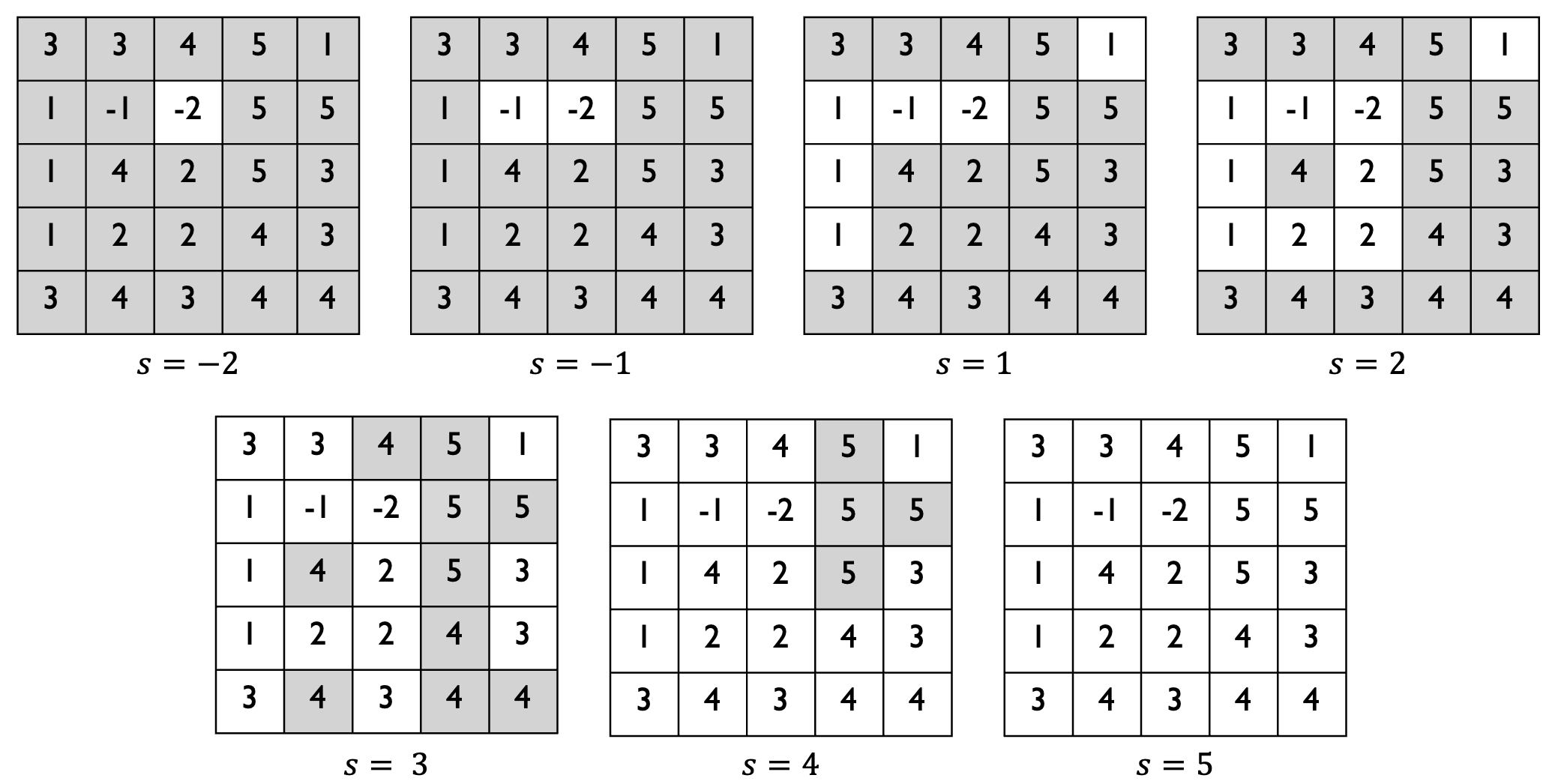}\\
        (a) Sub-level filtration. \\ \vspace{.2in}
      \includegraphics[width=0.30\textwidth]{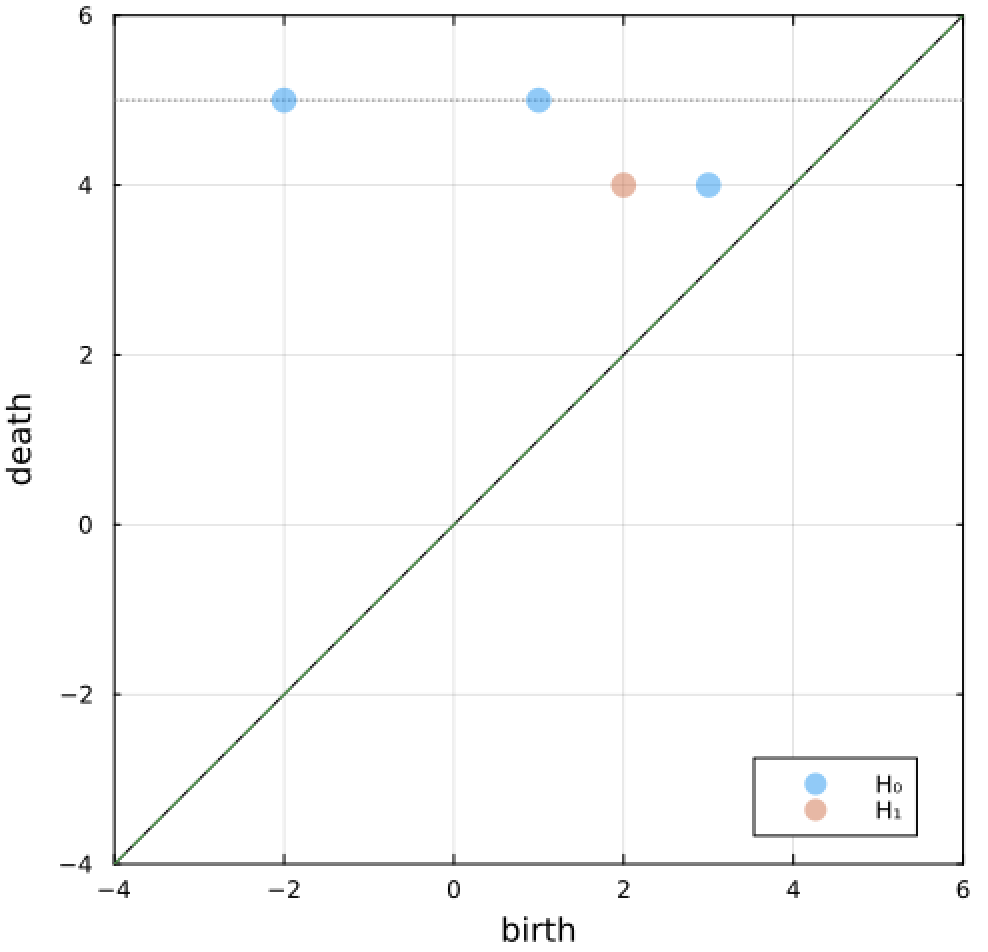} \\
      (b) Persistence diagram. \\
    \caption{Illustration of the Cubical filtration applied to a 2-dimensional grid image.}
    \label{fig:sublevel_Filtra}
\end{figure*}

To illustrate the concept of sublevel cubical filtrations, we present a toy example in Figure~\ref{fig:sublevel_Filtra}a. Here, the grid represents an image, where each square is assigned an artificial intensity value. As the threshold \( s \) increases, we can observe how connected components and loops emerge and evolve throughout the filtration. The filtration process begins at $s = -2$, where only a single pixel of intensity $-2$ at position $(2,3)$ is included. This lone pixel forms a single connected component with no loops, resulting in $\text{rank } H_0 = 1$ and $\text{rank } H_1 = 0$. At $s = -1$, the pixel with intensity $-1$ at $(2,2)$ joins and connects to the existing pixel, expanding the central component without creating any additional features. Thus, the homology ranks remain unchanged. By $s = 1$, all pixels with value $1$ or less are added, forming a larger structure on the left and including an isolated pixel at $(1,5)$. This increases the number of connected components to two, with $\text{rank } H_0 = 2$ and $\text{rank } H_1 = 0$. 

At $s = 2$, the inclusion of pixels with value $2$ expands the main structure further, and for the first time, a closed loop appears by surrounding a higher-value pixel, marking the birth of a one-dimensional hole. Meanwhile, the isolated pixel at $(1,5)$ persists, leaving the system with $\text{rank } H_0 = 2$ and $\text{rank } H_1 = 1$. When $s = 3$, pixels with value $3$ are added, enlarging the central component while maintaining the loop. Two new pixels at $(3,5)$ and $(4,5)$ also appear on the right as a separate component, and the isolated pixel remains, resulting in three connected components overall. The system now has $\text{rank } H_0 = 3$ and $\text{rank } H_1 = 1$. The picture changes at $s = 4$ when the pixel at $(3,2)$ fills in the previously existing loop, eliminating the one-dimensional hole. Additionally, right-side pixels connect with the central structure, reducing the total number of components to two. The only isolated pixel left is at $(1,5)$, so the ranks become $\text{rank } H_0 = 2$ and $\text{rank } H_1 = 0$. Finally, at $s = 5$, the last remaining pixels join in, including the isolated one in the top-right corner, merging everything into a single connected component. At this stage, the image is complete, with $\text{rank } H_0 = 1$ and $\text{rank } H_1 = 0$.\\

We summarize the evolution of topological features across all values of $s$ in $\mathcal{D}$ shown in Figure~\ref{fig:sublevel_Filtra}b. The blue points in $\mathcal{D}$ represent the birth and death coordinates \((\alpha_i, \alpha_j)\) of connected components (\(H_0\)). The persistence diagram shows two long-lived \(H_0\) features. One is born at \(s=-2\) and persists until \(s=5\), represented by the blue point \((-2, 5)\). Another persistent component appears at \(s=1\) and also survives until the end, reflected by the blue point at \((1, 5)\) (which is also illustrated in Figure~\ref{fig:sublevel_Filtra}a). The other connected component born at \( s = 3 \), however, merges quickly by \( s = 4 \) into the larger connected component that had appeared earlier on the left side. When two connected components are merged, the component with the larger birth-time is killed~\citep{kaji2020cubicalripsersoftwarecomputing}. This results in a short-lived \(H_0\) feature represented by the coordinates \((3, 4)\) near the diagonal. The  diagram also includes a  single orange point at \((2, 4)\) represents a one-dimensional hole (\(H_1\)) that was born at \( s = 2 \). At \( s = 4 \), the pixel in the center of that loop (with intensity 4) is added, filling in the hole and closing the loop. \\

\begin{figure}[!ht]
\centering
\includegraphics[width=0.99\textwidth]{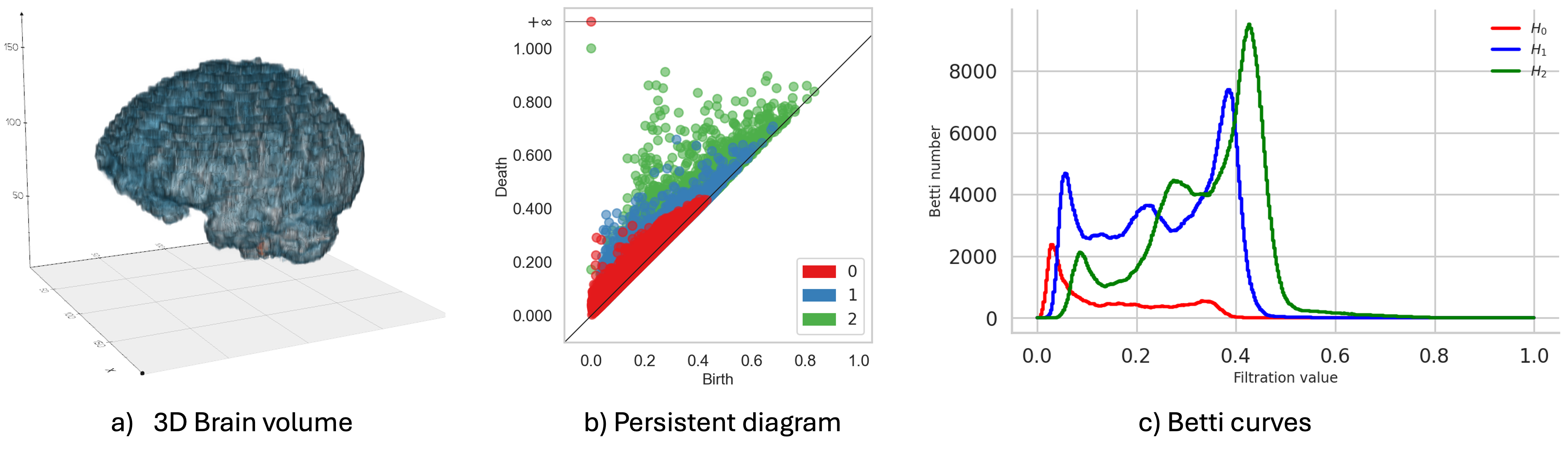}
  \caption{Topological feature extraction of brain MRI}
   \label{fig:MRI_to_Betti}
\end{figure}

We now apply TDA to a three-dimensional (3D) brain image and obtain topological summaries of the brain. Figure~\ref{fig:MRI_to_Betti} illustrates the extraction of topological features from the 3D brain volume using a cubical filtration. The image in panel (a) is transformed into a cubical complex across increasing filtration values, from which persistent homology is computed to evaluate topological structures. Panel (b) presents the resulting persistence diagram for dimensions $H_0$, $H_1$, and $H_2$, while panel (c) shows the corresponding Betti curves, summarizing the number of topological features across filtration levels.


\subsection{Functional representation of longitudinal topological summary of brain}

Let $I_{ij}$ denote the observed 3D brain image for subject $i=1,\ldots,n$ at visit $j=1,\ldots,m$. For each image $I_{ij}$, we construct a cubical filtration $\{\mathcal{K}_{ij,s}\}_{s\in\mathbb{R}}$ by thresholding the image intensity values over a common grid of filtration values $s_1,\dots,s_L$. For homology dimension $p\in\{0,1,2\}$, let $Y_{ij,p}(s_l), l=1,\ldots,L,$ denote the observed value of the $p$th Betti curve ($h_p$) for subject $i$ at visit $j$ and filtration value $s_l$. Equivalently, the full discretized Betti curve for subject $i$, visit $j$, and homology dimension $p$ can be represented by  
\begin{equation}
\label{eq:betti_curve_observed}
\mathbf{Y}_{ij,p}
=
\bigl\{
Y_{ij,p}(s_1),\ldots,Y_{ij,p}(s_L)
\bigr\}.
\end{equation}
Thus, we convert each imaging observation $I_{ij}$ into a count-valued functional summary of its topological structure over the filtration domain. In the longitudinal setting, subject $i$ contributes the repeated collection $\mathbf{Y}_{i1,p},\ldots,\mathbf{Y}_{im,p}$ which forms a set of longitudinal functional observation for homology dimension $p$. Variation in these functions across filtration values reflects changes in the number of connected components, loops, or voids, whereas variation across visits reflects longitudinal evolution in the topological structure of the underlying brain images. 

\section{Model specification}
\label{sec:3}

For each homology dimension $p$, we model the corresponding  Betti curves ${Y}_{ij,p}(s)$ as count-valued functional responses observed repeatedly for subject $i$ at visit $j$ over the domain $\mathcal{S}$. Repeated functional observations from the same subject are correlated across visits, while additional functional variation may arise at the individual-visit level. We therefore formulate a \textit{generalized multilevel functional regression model}, in which covariate effects vary smoothly over the domain $\mathcal{S}$. The subject-level latent functional process accounts for dependence among repeated observations, whereas the visit-level latent process captures additional visit-level functional variation not explained by the subject-level latent process. This formulation builds on functional additive mixed models for correlated and non-Gaussian functional responses, together with a multilevel representation of repeatedly observed functional data~\citep{di2009multilevel,scheipl2015functional, 10.1214/16-EJS1145}.

\subsection{Generalized multilevel functional regression model}


\label{sec:gen_multilevel_regression}
Let $Y_{ij,p}(s)$ denote the Betti functions for subject $i$ at visit $j$, in homology dimension $p$, evaluated at filtration value $s\in\mathcal{S}$ as defined in Eq.~\ref{eq:betti_curve_observed}. 
We consider subjects $i=1,\ldots,n$, each observed at visit $j=1,\ldots,m$. 
Our objective is to model the conditional mean of $Y_{ij,p}(s)$ while accounting for dependence among repeated functional observations within subject and additional variation arising at the individual-visit level. We model each homology dimension $p$ separately, because the corresponding Betti curves represent distinct topological features and may exhibit different distributional and functional characteristics. 
For each filtration value $s$, we assume that $Y_{ij,p}(s)$ follows a negative-binomial distribution with conditional mean $\mu_{ij,p}(s)$ and dispersion parameter $\theta_p > 0$. Under the adopted parameterization,
$$
\mathrm{E}\!\left\{Y_{ij,p}(s)\mid \cdot\right\}=\mu_{ij,p}(s) 
\qquad \text{and} \qquad
\mathrm{Var}\!\left\{Y_{ij,p}(s)\mid \cdot\right\}
=
\mu_{ij,p}(s)
+
\frac{\mu_{ij,p}(s)^2}{\theta_p}.
$$
This distributional specification accommodates overdispersion relative to the Poisson model while preserving the nonnegative count-valued nature of the Betti functions~\citep{lawless1987negative}. We model the conditional mean function $\mu_{ij,p}(s)$ using an additive functional predictor $\eta_{ij,p}(s)$ through the log link, 
\begin{equation}\nonumber
    \log \mu_{ij,p}(s)=\eta_{ij,p}(s),
\end{equation}

where, for each homology dimension $p$,
\begin{equation}
\label{eq:general_model}
\eta_{ij,p}(s)
=
\beta_{0,p}(s)
+
\mathbf{x}_{ij}^\top \boldsymbol{\beta_p}(s)
+
U_{i,p}(s)
+
W_{ij,p}(s).
\end{equation}

Here, $\beta_{0,p}(s)$ denotes the population-level intercept function. 
The vector $\mathbf{x}_{ij} = (x_{ij1},\dots,x_{ijr})^\top \in \mathbb{R}^r$ contains the scalar covariates associated with subject $i$ at visit $j$. Depending on the covariate, its value may vary over time or remain constant across repeated observations for the same subject~\cite{li2022fixed}. The corresponding coefficient functions are  $\boldsymbol{\beta_p}(s) = (\beta_{1,p}(s),\dots,\beta_{r,p}(s))^\top$, which measure how the associations between the scalar covariates and the log expected Betti count vary over the domain $s$. The latent process $U_{i,p}(s)$ models subject-level functional variation, thus allowing subjects to deviate from the population mean over the domain $s$. The process $W_{ij,p}(s)$ allows for functional variation specific to visit $j$ of subject $i$, and induces dependence among repeated visits from the same subject. Together, these two latent processes provide a multilevel representation of between-subject and within-subject functional variation~\citep{scheipl2015functional,10.1214/10-EJS575, 10.1214/14-AOAS748}. 
This formulation follows the generalized multilevel function-on-scalar regression framework of~\citep{goldsmith2015generalized}, together with the multilevel functional decomposition introduced by~\cite{di2009multilevel}, while using a negative-binomial distribution appropriate for the count-valued Betti responses.

\subsection{Multilevel functional principal component representation}
\label{sec:mul_fpca}

To identify the dominant modes of variation in the latent functional processes, we represent the subject- and visit-level processes using functional principal component decompositions. Functional principal component analysis provides a low-dimensional representation of a stochastic process through the eigenfunctions of its covariance operator, with the corresponding eigenvalues quantifying the variation along the principal directions~\cite{10.1214/009053606000000272}. For repeatedly observed functional data, this framework can be extended to separate the covariance structure arising at different levels of the sampling hierarchy~\cite{di2009multilevel,10.1214/10-EJS575,shou2015structured}.\\

For each homology dimension $p$, we assume that $U_{i,p}(s)$ and $W_{ij,p}(s)$ are mean-zero, square-integrable stochastic processes on $\mathcal{S}$, and that the subject-and visit-level processes are mutually uncorrelated. The subject- and visit-level covariance functions are defined as $\mathrm{C}_{U,p}(s,s') = \mathrm{Cov}\{U_{i,p}(s),U_{i,p}(s')\}$ and $\mathrm{C}_{W,p}(s,s') = \mathrm{Cov}\{W_{ij,p}(s),W_{ij,p}(s')\}$, respectively, for $s,s'\in \mathcal{S}$. The first covariance function captures between-subject functional variation, whereas the second captures within-subject variation between repeated visits. This decomposition provides a level-specific representation of the covariance structure of longitudinal functional observations~\cite{di2009multilevel}. Under standard regularity conditions~\cite{hsing2015theoretical}, the covariance operators $\mathrm{C}_{U,p}(s,s')$ and $\mathrm{C}_{W,p}(s,s')$ admit spectral decompositions  in terms of their eigenvalues and eigenfunctions, as 
$$
\mathrm{C}_{U,p}(s,s')
=
\sum_{k=1}^{\infty}
\lambda_{U,k,p}
{\phi}_{U,k,p}(s)
{\phi}_{U,k,p}(s')
\qquad \text{and} \qquad
\mathrm{C}_{W,p}(s,s')
=
\sum_{\ell=1}^{\infty}
\lambda_{W,\ell,p}
{\phi}_{W,\ell,p}(s)
{\phi}_{W,\ell,p}(s').
$$

Here $\lambda_{U,1,p} \geq \lambda_{U,2,p} \geq \cdots \geq 0$  and $\lambda_{W,1,p} \geq \lambda_{W,2,p} \geq \cdots \geq 0$ denote the ordered nonnegative subject- and visit-level eigenvalues, respectively, and $\phi_{U,k,p}(s)$ and $\phi_{W,\ell,p}(s)$ denote the corresponding eigenfunctions. Within each level, the eigenfunctions are orthonormal over the filtration domain
$\mathcal{S}$, such that $\int_{\mathcal{S}}
\phi_{U,k,p}(s)\phi_{U,k',p}(s)\,ds = \delta_{kk'},$
and $\int_{\mathcal{S}}
\phi_{W,\ell,p}(s)\phi_{W,\ell',p}(s)\,ds = \delta_{\ell\ell'},$ where $\delta$ denotes the Kronecker delta~\cite{di2009multilevel, 10.1214/009053606000000272}.\\ 

By the Karhunen--Lo\`eve theorem~\cite{hsing2015theoretical}, the two latent processes can be represented in terms of their functional principal components as \begin{equation}
\label{eq:KL}
U_{i,p}(s)
=
\sum_{k=1}^{\infty}
\xi_{ik,p}
\phi_{U,k,p}(s)
\qquad \text{and} \qquad
W_{ij,p}(s)
=
\sum_{\ell=1}^{\infty}
\zeta_{ij\ell,p}
\phi_{W,\ell,p}(s)
\end{equation}
where $\xi_{ik,p}$ and $\zeta_{ij\ell,p}$ are the subject- and visit-level functional principal component scores, respectively. Within each level, the scores have mean-zero and are uncorrelated across components, with $Var(\xi_{ik,p})=\lambda_{U,k,p}$ and $Var(\zeta_{ij\ell,p})=\lambda_{W,\ell,p}$. The eigenfunctions describe the dominant patterns of functional variation over the filtration domain, whereas the corresponding eigenvalues quantify the amount of variation explained by each pattern. The subject-level scores $\xi_{ik,p}$ quantify the contribution of the $k$th subject-level component for subject $i$, while the visit-level scores $\zeta_{ij\ell,p}$ quantify the contribution of the $\ell$th visit-level component for visit $j$ of subject $i$~\cite{10.1214/10-EJS575,shou2015structured}. 
In practice, the infinite Karhunen--Lo\`eve expansions are approximated by retaining finite numbers of components. Let $K_{U,p}$ and $K_{W,p}$ denote the numbers of subject- and visit-level components, respectively. The latent processes are then approximated by
\begin{equation}\nonumber
\label{eq:KL_finte}
U_{i,p}(s)\approx
\sum_{k=1}^{K_{U,p}}
\xi_{ik,p}
\phi_{U,k,p}(s)
\qquad \text{and} \qquad
W_{ij,p}(s)
\approx
\sum_{\ell=1}^{K_{W,p}}
\zeta_{ij\ell,p}
\phi_{W,\ell,p}(s).
\end{equation}

Substituting these truncated representations into our generalized multilevel functional regression model in Eq.~\ref{eq:general_model} gives
\begin{equation}
\label{eq:general_model_final}
\eta_{ij,p}(s)
=
\beta_{0,p}(s)
+
\mathbf{x}_{ij}^\top \boldsymbol{\beta_p}(s)
+
\sum_{k=1}^{K_{U,p}}
\xi_{ik,p}
\phi_{U,k,p}(s)
+
\sum_{\ell=1}^{K_{W,p}}
\zeta_{ij\ell,p}
\phi_{W,\ell,p}(s).
\end{equation}


\section{Estimation}
\label{sec:4}


We estimate our generalized multilevel functional regression model within a Bayesian framework, following the estimation approach of~\citet{goldsmith2015generalized}. For each homology dimension $p$, we represent the coefficient functions and the subject- and visit-level functional components using spline basis expansions. We then estimate all model components jointly, allowing uncertainty in both the fixed effects and the multilevel functional structure to be incorporated into the posterior distribution.

\subsection{Spline basis representation}

To obtain a finite-dimensional representation of our generalized multilevel functional model, we represent the coefficient functions and latent functional components using cubic B-spline basis expansions over $\mathcal{S}$. Basis expansions provide a flexible representation of smooth functions through a finite collection of basis coefficients and are widely used in functional data analysis~\citep{ramsay2005functional,crainiceanu2024functional}.

Let $\mathbf{b}(s)=\left(B_1(s),\ldots,B_{K_B}(s)\right)^\top$
denote a $K_B$-dimensional cubic B-spline basis defined on $\mathcal{S}$. For a generic coefficient function $g_{a,p}(s)$ in the fixed-effect component of the model, we use the representation $g_{a,p}(s)=\mathbf{b}(s)^\top\boldsymbol{\gamma}_{a,p},$
where $\boldsymbol{\gamma}_{a,p}=\left(\gamma_{a1,p},\ldots,\gamma_{aK_B,p}\right)^\top$ is the corresponding vector of spline coefficients. This representation is used for the intercept $\beta_{0,p}(s)$ and for each coefficient function contained in $\boldsymbol{\beta}_p(s)$.

We use the same spline basis to represent the functional components for the subject- and visit-level latent processes. During estimation, let $\widetilde{\phi}_{U,k,p}(s)$ and $\widetilde{\phi}_{W,\ell,p}(s)$ denote the pre-rotation subject- and visit-level component functions, respectively. These functions are represented as
$\widetilde{\phi}_{U,k,p}(s)=\mathbf{b}(s)^\top\boldsymbol{\gamma}_{U,k,p}$, $k=1,\ldots,K_{U,p},$ and $\widetilde{\phi}_{W,\ell,p}(s)=\mathbf{b}(s)^\top\boldsymbol{\gamma}_{W,\ell,p}$, $\ell=1,\ldots,K_{W,p},$
where $\boldsymbol{\gamma}_{U,k,p}$ and $\boldsymbol{\gamma}_{W,\ell,p}$ are the corresponding spline coefficient vectors. 
The pre-rotation subject- and visit-level component scores are denoted by $\widetilde{\xi}_{ik,p}$ and $\widetilde{\zeta}_{ij\ell,p}$, respectively. Accordingly, the latent processes used during estimation can be expressed as $$U_{i,p}(s)=\sum_{k=1}^{K_{U,p}}\widetilde{\xi}_{ik,p}\widetilde{\phi}_{U,k,p}(s)
\qquad \text{and} \qquad W_{ij,p}(s)=\sum_{\ell=1}^{K_{W,p}}\widetilde{\zeta}_{ij\ell,p}\widetilde{\phi}_{W,\ell,p}(s).
$$

We distinguish these pre-rotation component functions and scores from the orthonormal eigenfunctions and functional principal component scores defined in Section~\ref{sec:mul_fpca}, since orthonormality is not imposed directly during posterior sampling. The estimated component functions and scores are subsequently transformed using a post-estimation rotation to obtain an equivalent functional principal component representation with orthonormal eigenfunctions and corresponding component scores~\cite{goldsmith2015generalized}.

Since the Betti functions are evaluated on the common filtration grid $s_1,\ldots,s_L$, we collect the corresponding basis evaluations in the $L\times K_B$ matrix $\mathbf{B} = \left[ \mathbf{b}(s_1),\ldots,\mathbf{b}(s_L)\right]^\top.$
The values of a coefficient function over the observed filtration grid can then be expressed as $\mathbf{g}_{a,p}=\mathbf{B}\boldsymbol{\gamma}_{a,p}$,
with analogous representations for the subject- and visit-level component functions. 
We use a sufficiently rich B-spline basis to provide flexibility over the filtration domain, while smoothness is controlled through prior distributions on the spline coefficients. Regularization is introduced through smoothness-inducing priors rather than by restricting the basis dimension alone.

\subsection{Posterior estimation}
\label{sec:pos_est}

We impose smoothness on the coefficient functions and the pre-rotation functional components through Gaussian priors on their spline coefficients vectors. The prior precision is determined by a common penalty matrix that combines overall shrinkage with a second-order smoothness penalty. This formulation regularizes the spline coefficients while retaining sufficient flexibility to capture variation over the filtration domain.

Let $\mathbf{P}_0$ and $\mathbf{P}_2$ denote the zeroth- and second-order penalty components, respectively. Using the spline basis matrix $\mathbf{B}$, we define $\mathbf{P}_0 =\mathbf{B}^{\top}\mathbf{D}_0^{\top}\mathbf{D}_0\mathbf{B}$, and $\mathbf{P}_2 =\mathbf{B}^{\top}\mathbf{D}_2^{\top}\mathbf{D}_2\mathbf{B}$, where $\mathbf{D}_0$ is the zeroth-order difference operator and $\mathbf{D}_2$ is the second-order difference operator respectively. The penalty matrix is then defined as
$\mathbf{P}=\alpha \mathbf{P}_0+(1-\alpha)\mathbf{P}_2$. The value of $\alpha$ is set to $0.1$, placing greater weight on the second-order penalty to control functional roughness, while the smaller zeroth-order component provides additional shrinkage and ensures a proper positive-definite penalty matrix~\cite{goldsmith2015generalized}. 
For the spline coefficient vector $\boldsymbol{\gamma}_{a,p}$ corresponding to any ($a$th) population-level coefficient function, we specify the Gaussian prior, i.e., $\boldsymbol{\gamma}_{a,p} \mid \sigma_{a,p}^{2} \sim
N\left(\mathbf{0},\sigma_{a,p}^{2}\mathbf{P}^{-1}\right)$. Similarly, the spline coefficients for the pre-rotation subject- and visit-level component functions are assigned Gaussian priors, i.e., $\boldsymbol{\gamma}_{U,k,p} \mid \sigma_{U,k,p}^{2}\sim
N\left(\mathbf{0},\sigma_{U,k,p}^{2}\mathbf{P}^{-1}\right)$, $k=1,\ldots,K_{U,p}$, and $\boldsymbol{\gamma}_{W,\ell,p} \mid \sigma_{W,\ell,p}^{2}\sim
N\left(\mathbf{0},\sigma_{W,\ell,p}^{2}\mathbf{P}^{-1}\right)$, $\ell=1,\ldots,K_{W,p}$. Under this parameterization, the penalty matrix specifies the structure of the smoothness constraint, while the variance parameters determine the degree of regularization. Smaller smoothing variances induce smoother estimated functions, whereas larger values allow greater flexibility. \\

We assign independent inverse-gamma priors
to the smoothing variance parameters, $\sigma_{a,p}^{2}\sim
\operatorname{IG}(0.001,0.001)$, $\sigma_{U,k,p}^{2}\sim\operatorname{IG}(0.001,0.001)$, $\sigma_{W,\ell,p}^{2}\sim\operatorname{IG}(0.001,0.001)$.
For the latent component scores used during posterior sampling, we assign independent standard normal priors, $\widetilde{\boldsymbol{\xi}}_{i,p}\sim N_{K_{U,p}}\left(\mathbf{0},\mathbf{I}_{K_{U,p}}\right)$, where $i=1,..,n, k=1,...,K_{U,p}$ and $\widetilde{\boldsymbol{\zeta}}_{ij,p}\sim N_{K_{W,p}} \left( \mathbf{0},\mathbf{I}_{K_{W,p}}
\right)$, where $i=1,...n, j=1,...m, \ell=1,...,K_{W,p}$. These priors are imposed on the pre-rotation scores. The conventional functional principal component scores and their corresponding variances are obtained after the post-estimation rotation, as described in section~\ref{sec:rotation}. Finally, for the negative-binomial dispersion parameter, introduced in the response model, we specify the prior $\theta_p \sim \operatorname{Lognormal}\{\log(10),\,1.5\}$.\\  

We obtain posterior samples for all model parameters jointly using Stan, implemented through the RStan interface~\cite{JSSv076i01}. In particular, posterior sampling is performed using the No-U-Turn Sampler, an adaptive form of Hamiltonian Monte Carlo that is well suited for Bayesian models with high-dimensional and correlated parameter structures~\cite{JMLR:v15:hoffman14a,goldsmith2015generalized}. 
Two independent Markov chains are run for $3000$ iterations each, with the first $1000$ iterations of each chain discarded as warm-up. 
The spline basis dimension is fixed at $K_B = 10$. For the multilevel functional representation, the number of functional principal components $K_{U,p}$ and $K_{W,p}$ are required to be specified before model fitting. We assess sensitivity to this choice by fitting models with different components at each level.

We evaluate convergence and sampling performance using trace plots, posterior density plots, the potential scale reduction statistic $\widehat{R}$, and effective sample sizes. Values of $\widehat{R}$ close to one and sufficiently large effective sample sizes and stable mixing across chains are taken as evidence of satisfactory convergence~\citep{e2c4473124534b8e930c11464ae1b1dd}. Posterior inference for the regression coefficient functions is based on posterior means and pointwise $95\%$ credible intervals obtained from the corresponding posterior quantiles. 

\subsection{Rotation}
\label{sec:rotation}

The Bayesian estimation procedure described in Section~\ref{sec:pos_est} provides estimates of the subject- and visit-level component functions together with their corresponding scores. However, the estimated component functions are not constrained to be orthonormal during posterior sampling, and the ordering of the resulting components is not uniquely determined. We therefore apply a post-estimation rotation separately at the subject- and visit-level to obtain the conventional functional principal component representation with orthonormal eigenfunctions and ordered component variances.\\

For the subject level, let $\widetilde{\boldsymbol{\Phi}}_{U,p}$ denote the matrix of estimated component functions evaluated over the filtration grid. The singular value decomposition of $\widetilde{\boldsymbol{\Phi}}_{U,p}$ is given by $\widetilde{\boldsymbol{\Phi}}_{U,p} = \mathbf{Q}_{U,p}\mathbf{D}_{U,p}\mathbf{V}_{U,p}^{\top}$, where the columns of $\mathbf{Q}_{U,p}$ and $\mathbf{V}_{U,p}$ are the left and right singular vectors, respectively, and are orthonormal, while $\mathbf{D}_{U,p}$ is a diagonal matrix containing the nonnegative singular values in decreasing order. Under this decomposition, the subject-level component functions and their corresponding scores are rotated to obtain equivalent representation with orthonormal component functions while preserving the reconstructed subject-level latent process.\\

Under the independent standard normal priors assigned to the pre-rotation scores, the variances of the rotated scores are determined by the squared singular values. Accordingly, the subject-level component variances are $\lambda_{U,k,p} = d_{U,k,p}^{2}$, for $k = 1,\ldots,K_{U,p}$, and are ordered as $\lambda_{U,1,p}\geq \lambda_{U,2,p}\geq\cdots\geq \lambda_{U,K_{U,p},p}\geq0$. The same rotation is applied to the visit-level components, with component variances $\lambda_{W,\ell,p} = d_{W,\ell,p}^{2}$, for $\ell = 1,\ldots,K_{W,p}$, ordered as $\lambda_{W,1,p}\geq \lambda_{W,2,p}\geq\cdots\geq \lambda_{W,K_{W,p},p}\geq0$. The rotation is carried out within each posterior iteration. Prior to computing posterior summaries, the signs of the rotated eigenfunctions are aligned across posterior draws to ensure a consistent orientation.\\

\section{Application to Alzheimer’s disease progression}
\label{sec:5}

\subsection{Data and variables}
\label{sec:Data}
We use data from the \textit{Open Access Series of Imaging Studies (OASIS-2)}~\citep{OASIS2Website}, a publicly available longitudinal neuroimaging dataset developed to support research on aging and Alzheimer's disease progression. The OASIS-2 release contains longitudinal T1-weighted structural MRI scans from 150 older adults aged 60--96 years, collected over multiple visits separated by at least one year, for a total of 373 imaging sessions. For this study, we include participants with three MRI visits and focus on those with dementia, including participants who converted from nondemented to demented during follow-up. This selection yields 22 participants, resulting in a total of 66 MRI examinations across the three visits. Demographic and clinical characteristics of the participants included in the analysis are provided in Table 2 of the Appendix.\\

For each of the 66 retained MRI examinations, we quantify brain topology using Betti curves. The original T1-weighted MRI scans are provided in NIfTI format and are preprocessed using the Swiss Stripper extension in 3D Slicer to remove the skull and other nonbrain structures, thereby restricting the analysis to the brain region~\cite{fedorov2012slicer}. The resulting skull-stripped MRI dataset for each examination contains the complete three-dimensional brain volume represented by voxels. For each 3D image $I_{ij}$, we construct a cubical filtration over the common filtration grid $s_1,\ldots,s_L$ and compute the corresponding Betti curves $Y_{ij,p}(s)$ for homology dimensions $p=0,1,2$, as described in Section~\ref{sec:2}. We use these resulting count-valued Betti curves as the functional responses $Y_{ij,p}(s)$ in the subsequent analysis. \\

\begin{figure}[htbp]
    \centering

    \begin{minipage}[t]{0.32\textwidth}
        \centering
        \includegraphics[width=\linewidth]{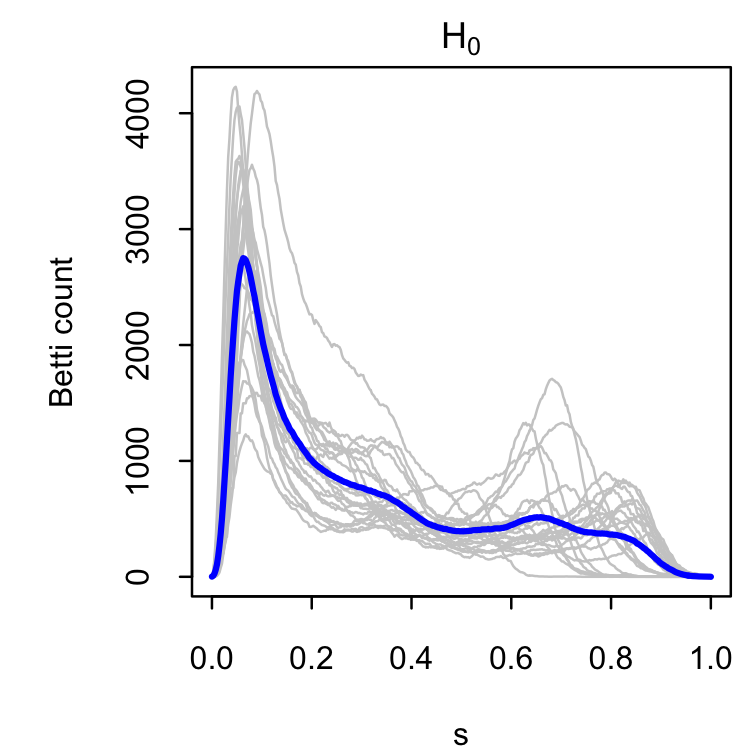}
    \end{minipage}
    \hspace{-0.0025\textwidth}
    \begin{minipage}[t]{0.32\textwidth}
        \centering
        \includegraphics[width=\linewidth]{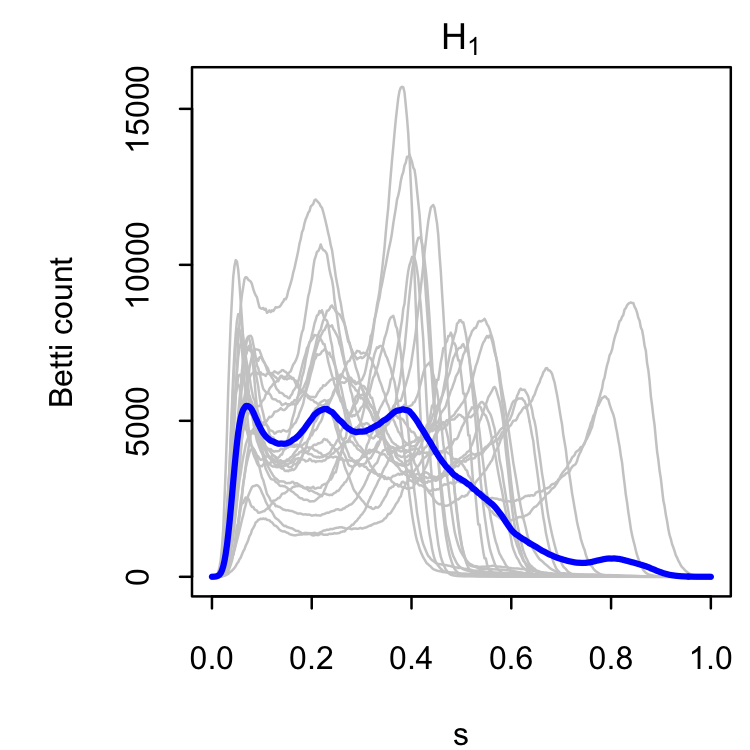}
    \end{minipage}
    \hspace{-0.0025\textwidth}
    \begin{minipage}[t]{0.32\textwidth}
        \centering
        \includegraphics[width=\linewidth]{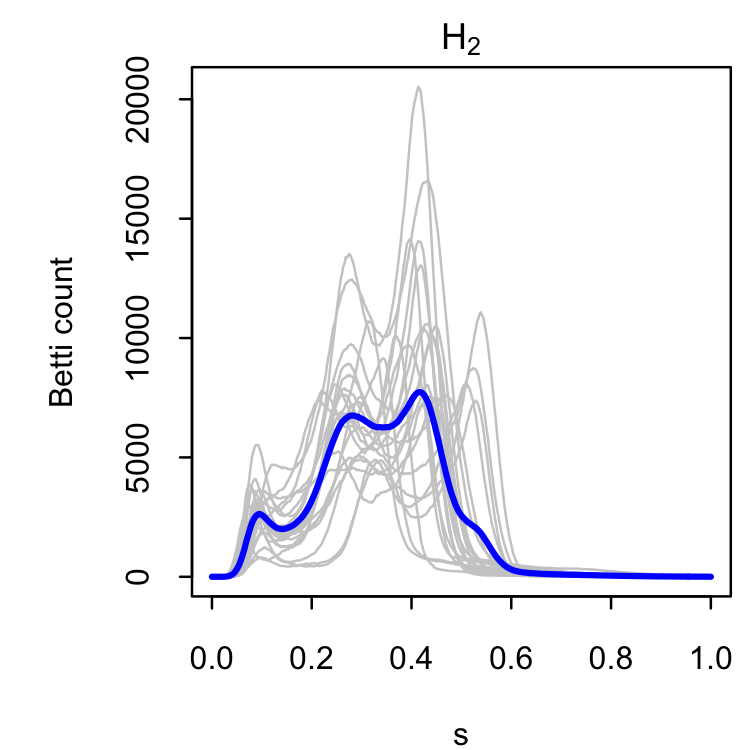}
    \end{minipage}

    \caption{Observed $H_0$, $H_1$, $H_2$ Betti curves at the first visit. Gray lines represent individual subject curves, and the blue line represents the mean Betti curve across subjects.}
    \label{fig:visit1_betti}
\end{figure}

Figure~\ref{fig:visit1_betti} displays the observed $H_0$, $H_1$, and $H_2$ Betti curves at the first visit, illustrating the variation in topological structure across participants over the domain $s$. We find that the $H_0$ curves show a pronounced early peak followed by a gradual decline, with between-subject variation becoming more evident over the middle and higher values of $s$. In contrast, the \(H_1\) curves exhibit considerably greater between-subject variability, with pronounced differences in the shape and magnitude of their multiple peaks. 
The $H_2$ curves are concentrated primarily in the middle range of $s$, where the dominant peaks occur, and show less between-subject variability than $H_1$ curves. The $H_0$, $H_1$, and $H_2$ Betti curves for the second and third visits are presented in Figures~9 and 10 of the Appendix, respectively. We observe similar patterns across the second and third visits, consistent with those shown in Figure~\ref{fig:visit1_betti}.

\subsection{Model and Results}

To examine how the longitudinal Betti curves vary with demographic and clinical covariates while accounting for within-subject dependence arising from repeated measurements, we fit the negative-binomial generalized multilevel functional regression model defined in Eq.~\ref{eq:general_model_final}. 
For each homology dimension $p \in \{0,1,2\}$, the model is specified as
\begin{align}
\eta_{ij,p}(s)
={}&
\beta_{0,p}(s)
+\beta_{1,p}(s)\mathrm{Age}_{i}
+\beta_{2,p}(s)\mathrm{Male}_{i}
+\beta_{3,p}(s)\mathrm{Time}_{ij} \nonumber\\
&+\beta_{4,p}(s)I\{\mathrm{CDR}_{ij}=0.5\}
+\beta_{5,p}(s)I\{\mathrm{CDR}_{ij}=1\} \nonumber\\
&+\sum_{k=1}^{3}
\xi_{ik,p}\phi_{U,k,p}(s)
+\sum_{\ell=1}^{3}
\zeta_{ij\ell,p}\phi_{W,\ell,p}(s),
\label{eq:app_model}
\end{align}

where $\mathrm{Age}_i$ denotes baseline age, defined as the participant's age at the first MRI visit and centered at 75 years. $\mathrm{Male}_i$ is an indicator for gender, with female as the reference category. $\mathrm{Time}_{ij}$ denotes the elapsed time, in years, from subject $i$'s first MRI visit to visit $j$, with the first visit corresponding to time zero. CDR is included as a categorical covariate, with $\mathrm{CDR}=0$ as the reference category and indicator variables for $\mathrm{CDR}=0.5$ and $\mathrm{CDR}=1$, corresponding to very mild and mild dementia, respectively. Accordingly, $\beta_{1,p}(s)$, $\beta_{2,p}(s)$, and $\beta_{3,p}(s)$  represent the associations of age, gender, and follow-up time with the log expected Betti curve, while $\beta_{4,p}(s)$, and $\beta_{5,p}(s)$ describe the differences associated with, $\mathrm{CDR}=0.5$, and $\mathrm{CDR}=1$ relative to $\mathrm{CDR}=0$. We choose $K_{U,p} = 3$ and $K_{W,p} = 3$ based on the sensitivity analysis as described in Section~\ref{sec:pos_est}.
\\

\begin{figure}[htbp]
    \centering

    \begin{minipage}[t]{0.33\textwidth}
        \centering
        \includegraphics[width=\linewidth]{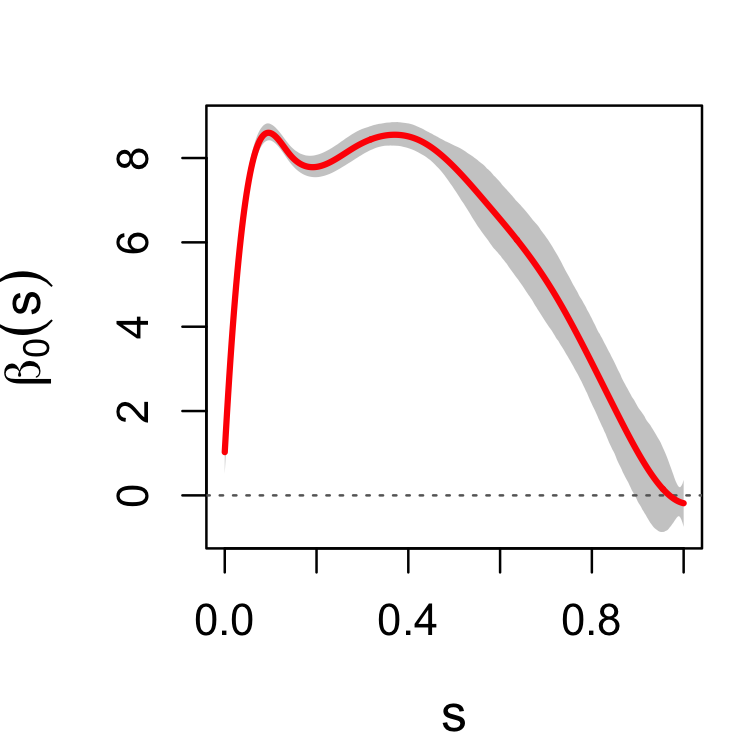}
    \end{minipage}
    \hfill
    \begin{minipage}[t]{0.33\textwidth}
        \centering
        \includegraphics[width=\linewidth]{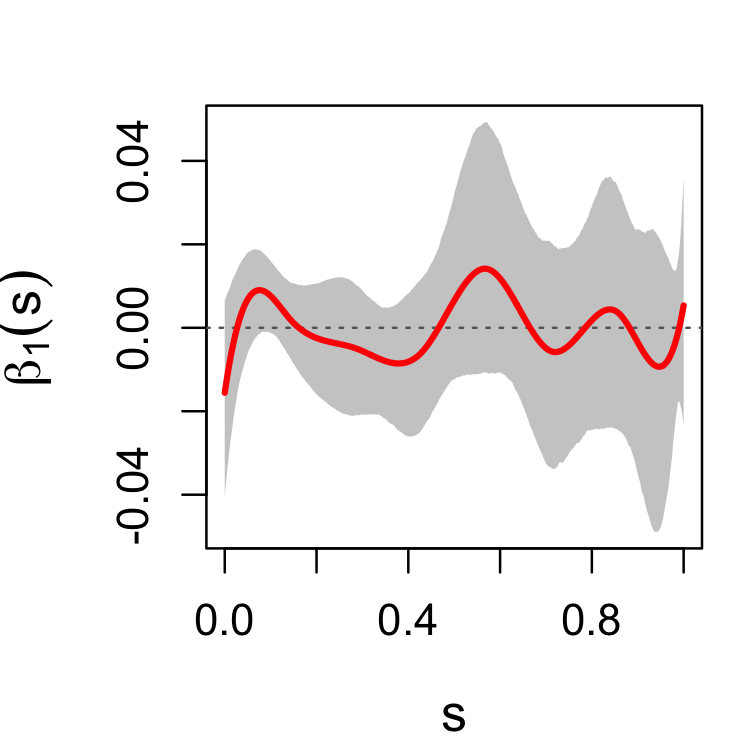}
    \end{minipage}
    \hfill
    \begin{minipage}[t]{0.33\textwidth}
        \centering
        \includegraphics[width=\linewidth]{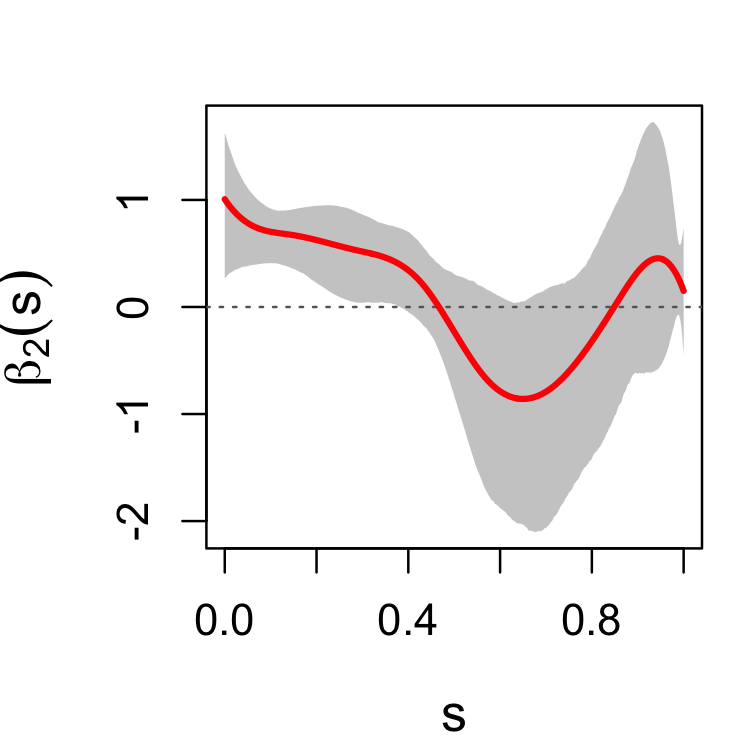}
    \end{minipage}
    \hfill
    \begin{minipage}[t]{0.33\textwidth}
        \centering
        \includegraphics[width=\linewidth]{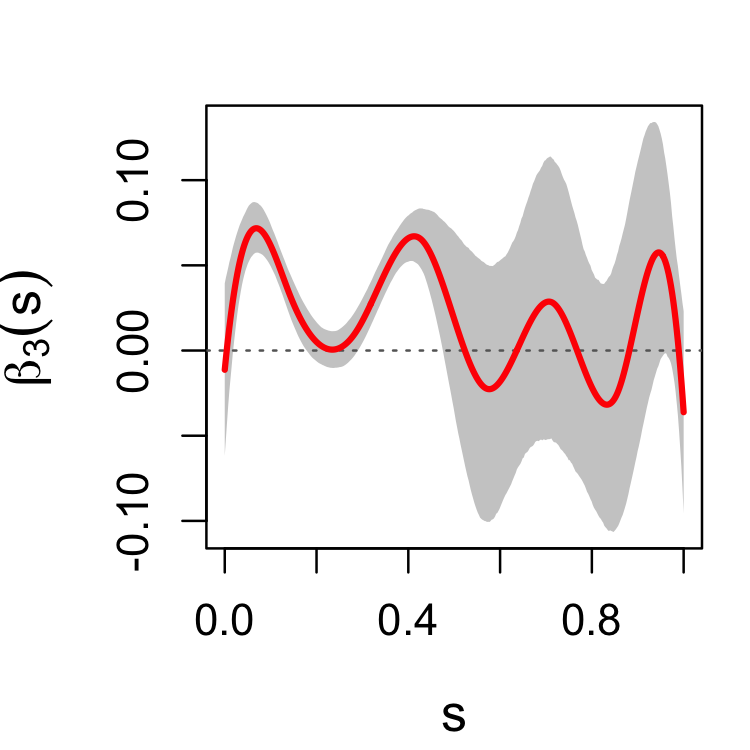}
    \end{minipage}
    \hfill
    \begin{minipage}[t]{0.33\textwidth}
        \centering
        \includegraphics[width=\linewidth]{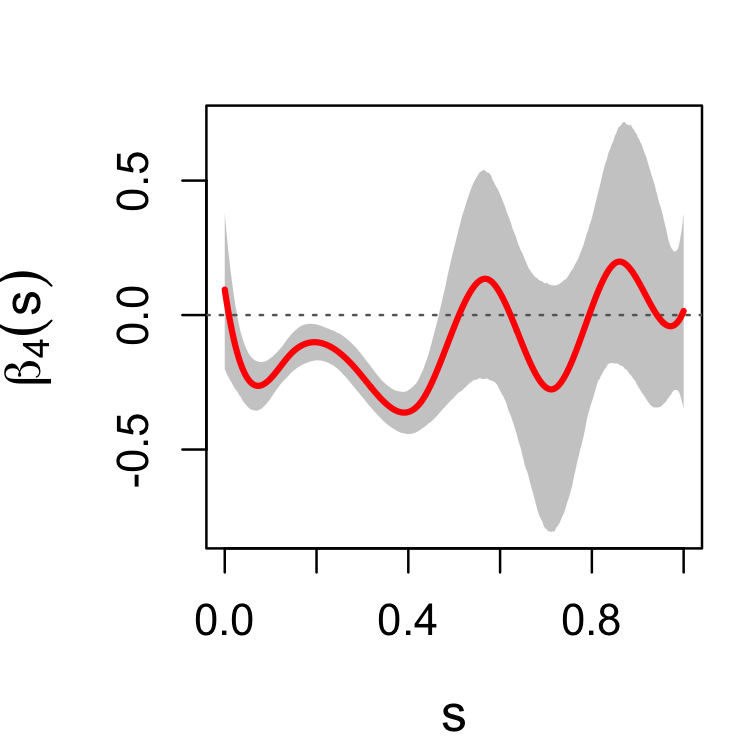}
    \end{minipage}
    \hfill
    \begin{minipage}[t]{0.33\textwidth}
        \centering
        \includegraphics[width=\linewidth]{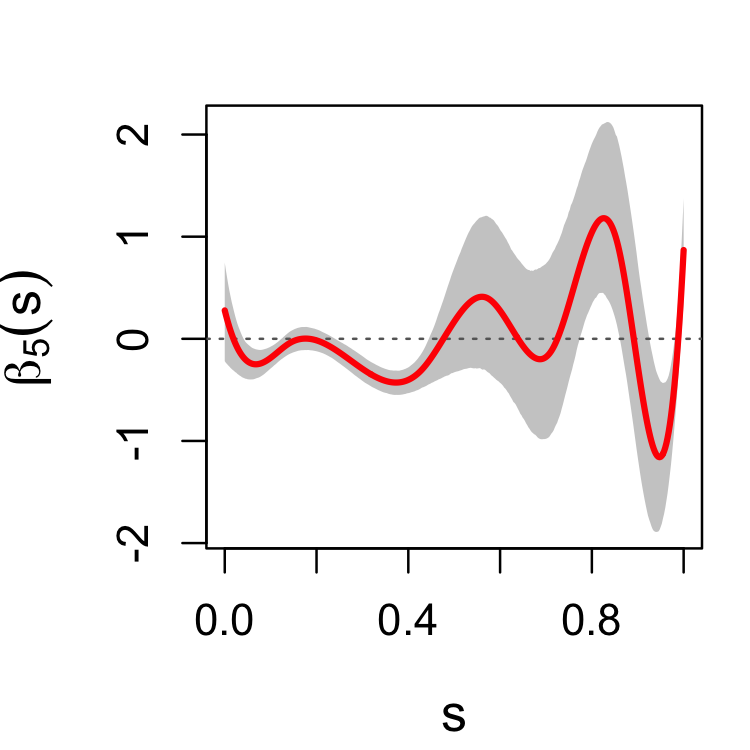}
    \end{minipage}

    \caption{Posterior mean coefficient functions and corresponding pointwise 95\% credible intervals for $H_1$.}
    \label{fig:coef_h1}
\end{figure}

We first examine the estimated fixed-effect coefficient functions to assess how the demographic and clinical covariates are related to the Betti curves over $s$. Figures~\ref{fig:coef_h1} and~\ref{fig:coef_h2} present the posterior mean coefficient functions and corresponding pointwise 95\% credible intervals for $H_1$ and $H_2$, respectively (corresponding $H_0$ results are presented in the Appendix Figure 11). The coefficient functions are shown on the log-mean scale. In particular, the intercept function $\beta_{0,p}(s)$ represents the log expected Betti count for a 75-year-old female at the first MRI visit with $\mathrm{CDR}=0$, when the subject- and visit-level latent processes are set to zero. Consequently, $\exp\{\beta_{0,p}(s)\}$ gives the corresponding expected Betti curve on the original count scale. \\
For $H_1$, the intercept function $\beta_{0}(s)$ shows a sharp early peak with a broader second peak and then gradually declines towards zero at  high filtration values. The estimated baseline-age coefficient $\beta_{1}(s)$ remains close to zero across most values of $s$, with the 95\% credible interval generally including zero, providing limited evidence of baseline age effect. The gender coefficient $\beta_{2}(s)$ is more pronounced, particularly at lower values of $s$, suggesting higher expected $H_1$ Betti counts for males than females over this region. The follow-up-time coefficient $\beta_{3}(s)$ and CDR coefficients ($\beta_{4}(s)$, and $\beta_{5}(s)$) vary around zero across $s$, with wider credible intervals at higher values of $s$, indicating greater uncertainty in the these estimates.
\begin{figure}[htbp]
    \centering
    \begin{minipage}[t]{0.33\textwidth}
        \centering
        \includegraphics[width=\linewidth]{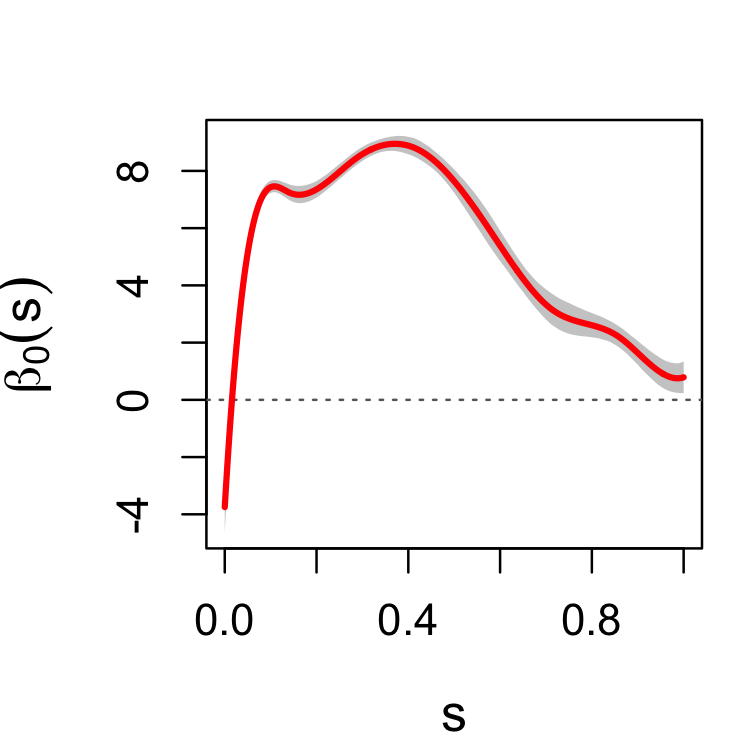}
    \end{minipage}
    \hfill
    \begin{minipage}[t]{0.33\textwidth}
        \centering
        \includegraphics[width=\linewidth]{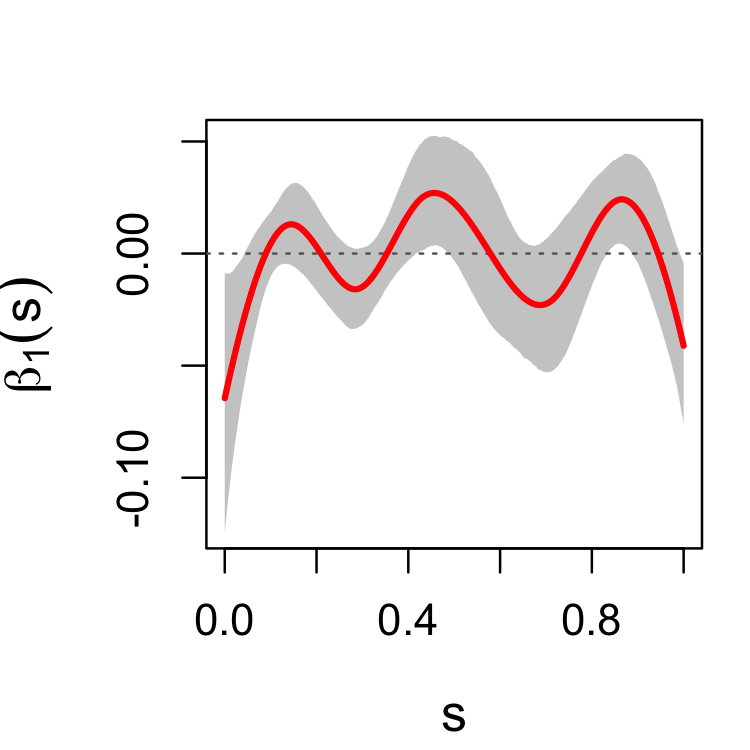}
    \end{minipage}
    \hfill
    \begin{minipage}[t]{0.33\textwidth}
        \centering
        \includegraphics[width=\linewidth]{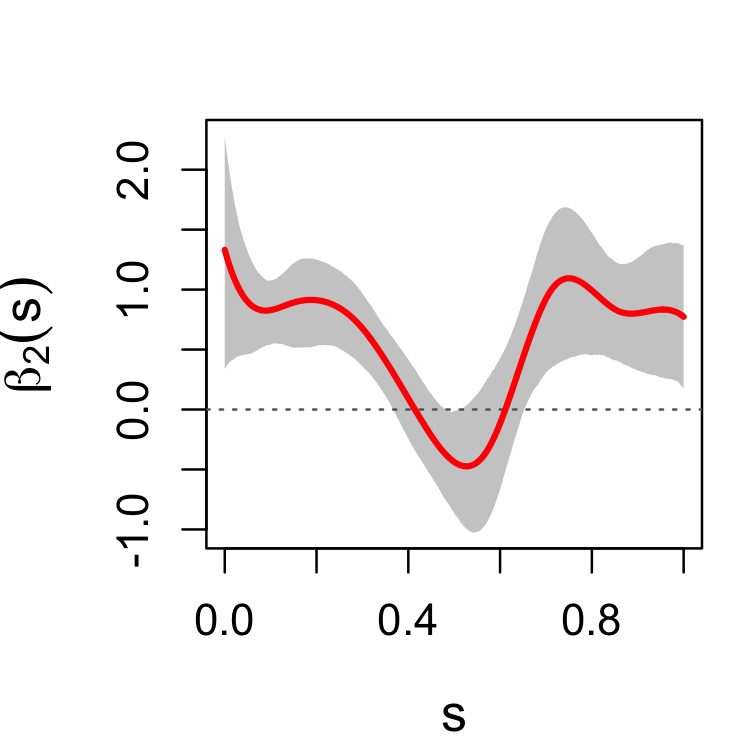}
    \end{minipage}
    \hfill
    \begin{minipage}[t]{0.33\textwidth}
        \centering
        \includegraphics[width=\linewidth]{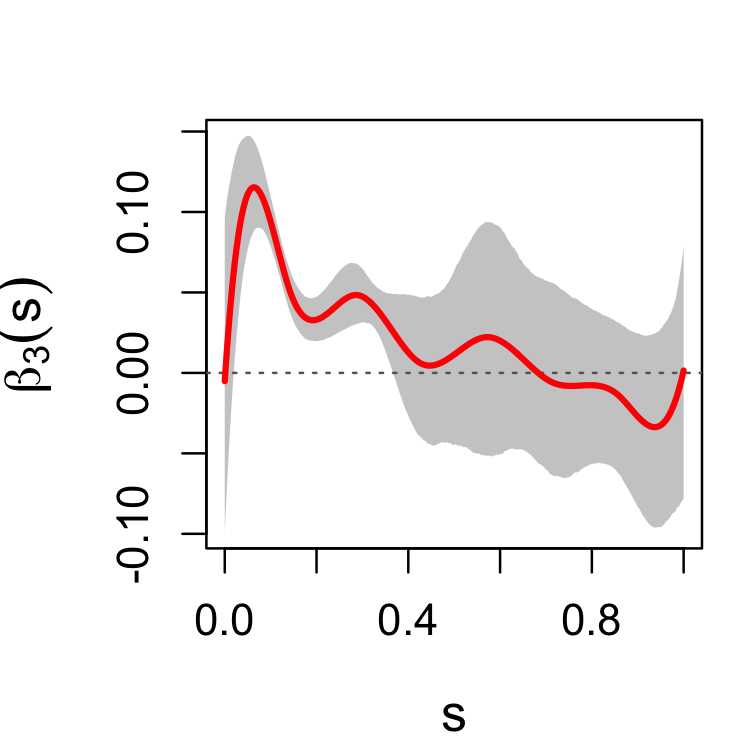}
    \end{minipage}
    \hfill
    \begin{minipage}[t]{0.33\textwidth}
        \centering
        \includegraphics[width=\linewidth]{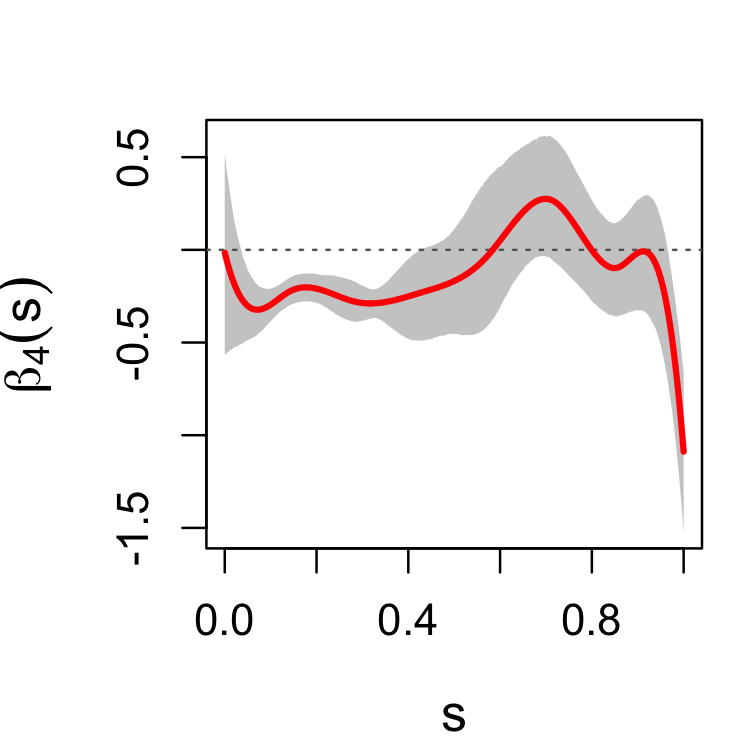}
    \end{minipage}
    \hfill
    \begin{minipage}[t]{0.33\textwidth}
        \centering
        \includegraphics[width=\linewidth]{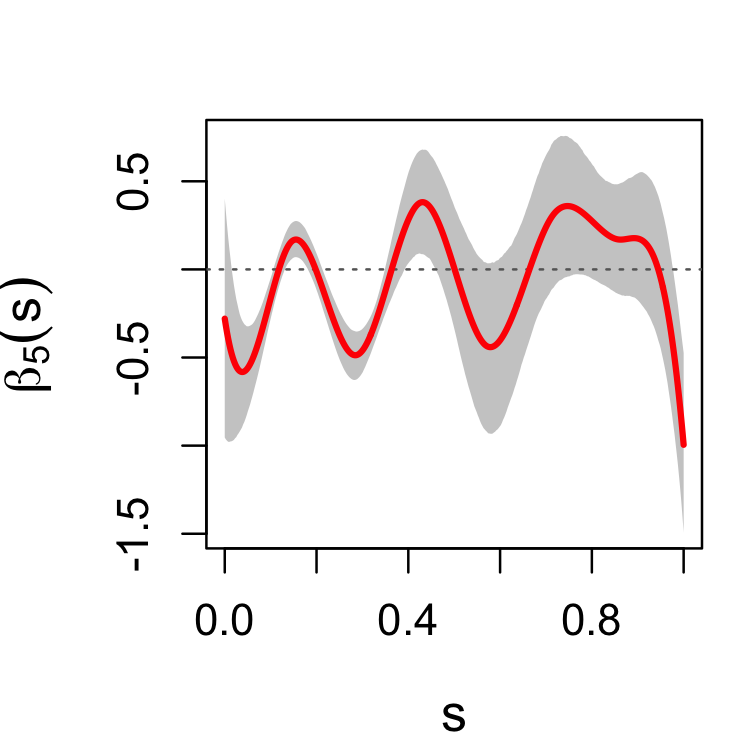}
    \end{minipage}
    \caption{Posterior mean coefficient functions and corresponding pointwise 95\% credible intervals for $H_2$.}
    \label{fig:coef_h2}
\end{figure}
For $H_2$, the estimated coefficient patterns differ from those observed for $H_1$. The intercept function increases rapidly at lower filtration values, reaches its highest values near the center of the filtration domain, and then gradually declines.
The baseline-age coefficient remains close to zero, whereas the gender coefficient shows more pronounced variation over $s$, with a positive effect over the earlier and later portions of the curve and a negative effect in the middle range. The follow-up-time coefficient is positive at lower values of $s$ and approaches zero thereafter. The CDR coefficients also vary over $s$, although their credible intervals are relatively wide, indicating greater uncertainty.\\

We conduct posterior convergence diagnostics for all three fitted models. In particular, we evaluate trace plots, posterior density estimates, $\widehat{R}$ statistics, and effective sample sizes. The results are presented in Appendix Section~A.4 (Table 3 and Figures 12– 14). The diagnostic results indicate convergence for all three fitted models, with well-mixed trace plots, consistent posterior density estimates across chains, $\widehat{R}$ values close to 1, and adequate effective sample sizes.\\

\begin{figure}[htbp]
    \centering
    \includegraphics[width=0.8\linewidth]{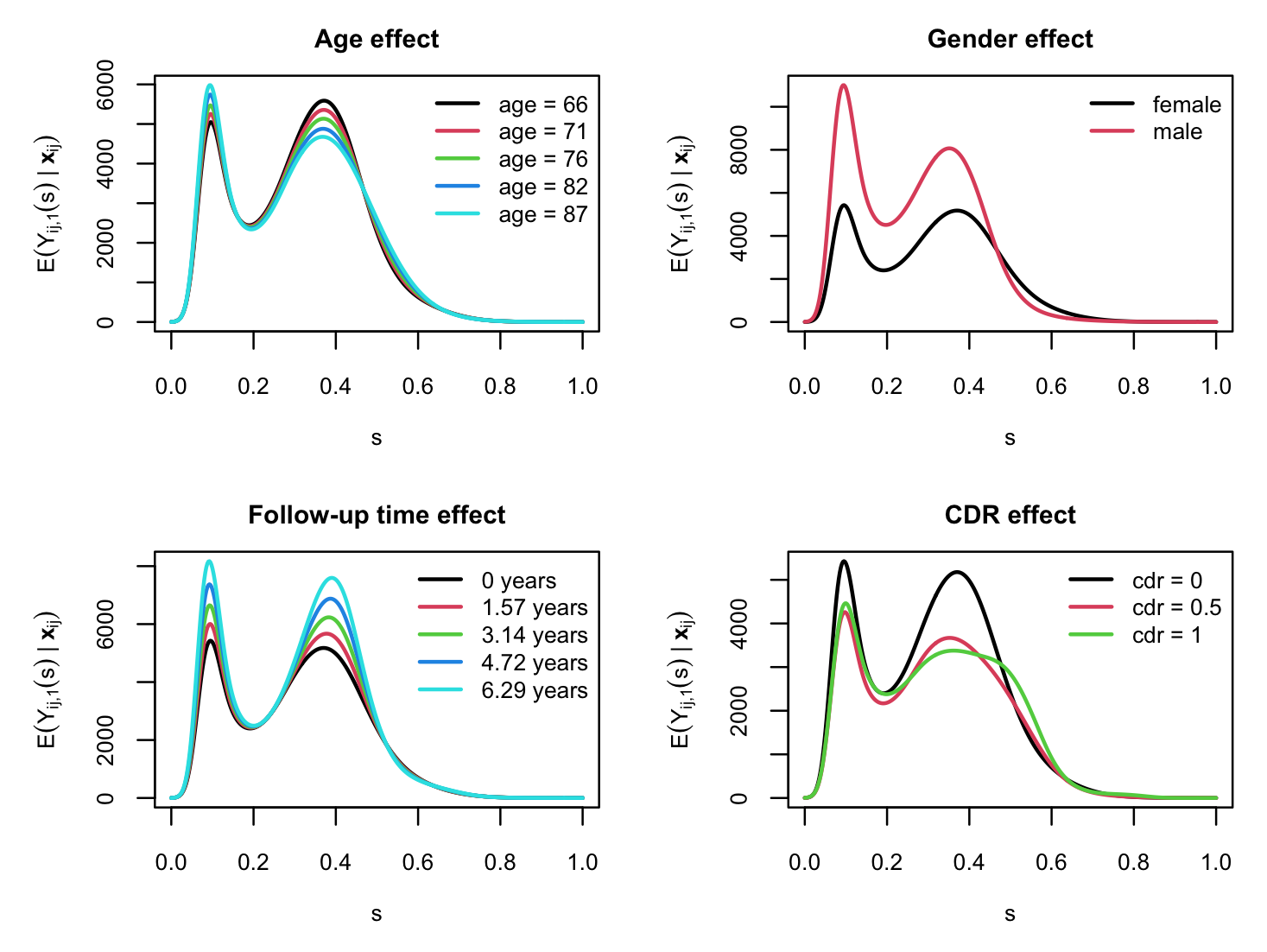}
    \caption{Estimated population-level $H_1$ Betti curves on the original count scale for selected values of baseline age, gender, follow-up time, and CDR. Within each panel, the remaining covariates are held at their reference values, and the subject- and visit-level latent effects are set to zero.}
    \label{fig:coef_h1_effect}
\end{figure}

\begin{figure}[htbp]
    \centering
    \includegraphics[width=0.8\linewidth]{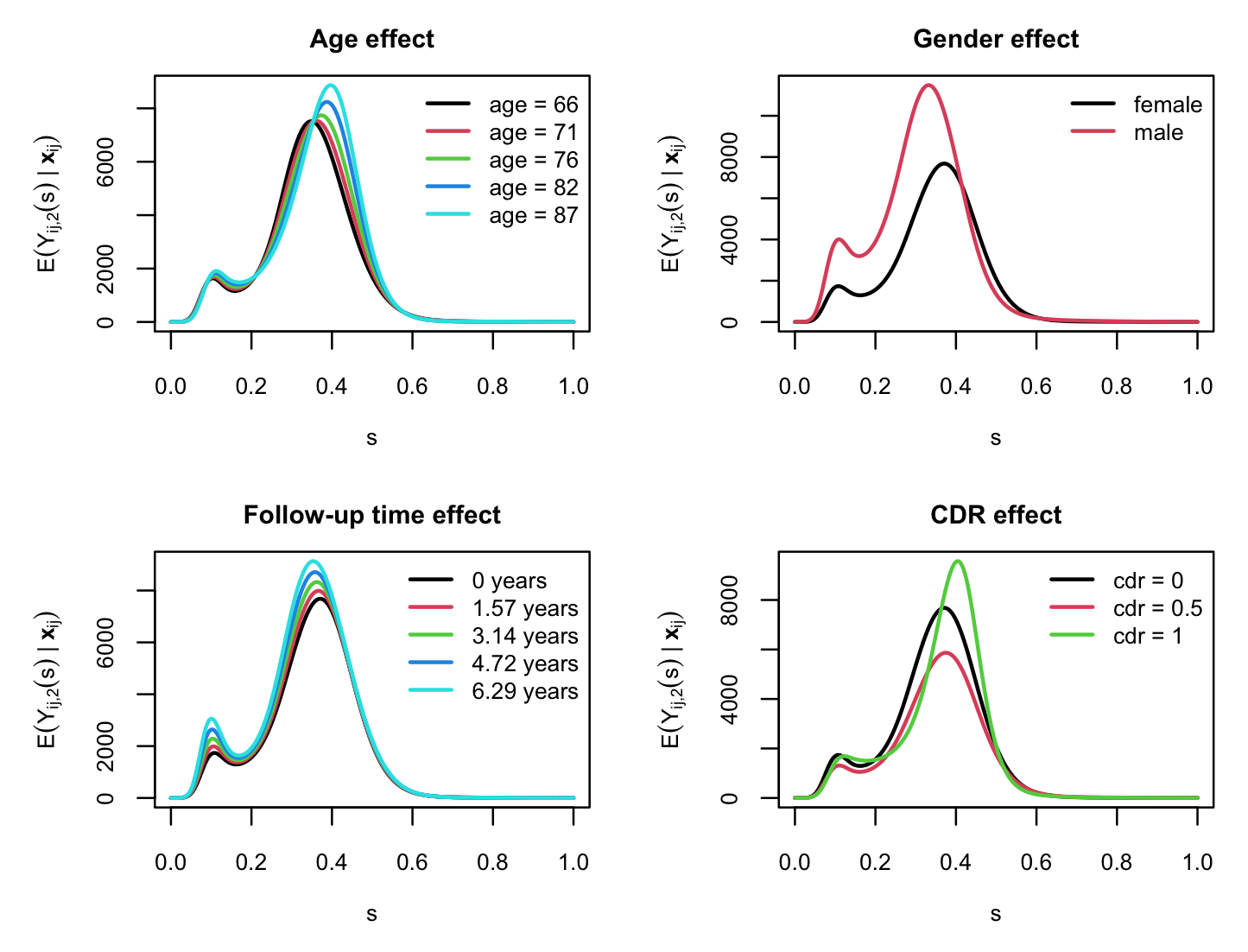}
    \caption{Estimated population-level $H_2$ Betti curves on the original count scale for selected values of baseline age, gender, follow-up time, and CDR. Within each panel, the remaining covariates are held at their reference values, and the subject- and visit-level latent effects are set to zero.}
    \label{fig:coef_h2_effect}
\end{figure}

To provide a more interpretable representation of the estimated fixed effect on the original count scale, we evaluate expected $H_1$ and $H_2$ Betti curves at selected covariate values (See Figures~\ref{fig:coef_h1_effect} and~\ref{fig:coef_h2_effect}). For each comparison, the remaining covariates are held at their reference values, and the subject- and visit-level latent effects are set to zero.
The response-scale curves in Figure~\ref{fig:coef_h1_effect} reveal distinct covariate-related changes in the $H_1$ topological structure. 
Increasing baseline age corresponds to a higher first peak and a lower second peak, indicating that the age effect differs across $s$ rather than producing a uniform increase or decrease in the expected number of one-dimensional topological features. The gender comparison shows substantially higher expected $H_1$ Betti counts for males around both major peaks. Longer follow-up time is also associated with progressively higher counts around the two peaks, while the overall locations of these features remain similar. 
Differences according to
CDR are most evident around the peak regions: both $\mathrm{CDR}=0.5$ and
$\mathrm{CDR}=1$ generally show lower expected counts than $\mathrm{CDR}=0$, with the $\mathrm{CDR}=1$ curve also showing a broader second peak extending toward higher values of $s$.\\

The corresponding $H_2$ curves in Figure~\ref{fig:coef_h2_effect} also exhibit several distinct patterns across the domain $s$. Increasing baseline age is associated with a progressively higher dominant peak and a modest shift toward larger values of $s$, while the smaller early peak remains relatively stable. Males show higher expected $H_2$ Betti counts than females around both the early and dominant peaks. Longer follow-up is also associated with progressively higher expected counts, particularly around the dominant peak. The CDR profiles show a different pattern: relative to $\mathrm{CDR}=0$, $\mathrm{CDR}=0.5$ is associated with a lower
dominant peak, whereas $\mathrm{CDR}=1$ is associated with a higher peak occurring at a slightly larger value of $s$. 
Similar distinct covariate-related changes in the $H_0$ curves are observed in the Appendix Figure~15. 


These results demonstrate that the topological summaries reveal distinct covariate-associated patterns that vary across the filtration domain. In particular, changes in age, gender, follow-up time, and dementia severity are reflected in the overall number of topological features as well as in the relative prominence and shape of different regions of the Betti curves.\\


We now evaluate the remaining multilevel functional variation after accounting for the population-level covariate effects based on the last two components of Eq.~\ref{eq:app_model}. 
Table~\ref{tab:variance_proportions} summarizes the posterior mean proportion of functional variation attributed to the subject- and visit-level latent processes, together with the corresponding 95\% credible intervals. We find that visit-level variation accounts for the majority of the variability in all three homology dimensions, contributing approximately 97.03\%, 91.23\%, and 81.78\% for $H_0$, $H_1$, and $H_2$, respectively. However, notice that the contribution of subject-level variation increases with higher homology dimensions, from 2.97\% for $H_0$ to 8.78\% for $H_1$ and 18.26\% for $H_2$. Thus, although within-subject variation across visits remains dominant, between-subject differences become increasingly prominent in the higher-dimensional topological features.

\begin{table}[htbp]
\centering
\caption{Proportion of multilevel functional variation attributed to the subject and visit levels.}
\label{tab:variance_proportions}

\renewcommand{\arraystretch}{1.35}
\setlength{\tabcolsep}{14pt}

\begin{tabular}{lcc}
\hline
\textbf{Homology dimension} &
\textbf{Subject level, \% (95\% CI)} &
\textbf{Visit level, \% (95\% CI)} \\
\hline
$H_0$ & 2.97 (1.41, 6.22) & 97.03 (93.78, 98.60) \\
$H_1$ & 8.78 (4.33, 17.07) & 91.22 (82.93, 95.67) \\
$H_2$ & 18.26 (11.42, 28.06) & 81.78 (71.94, 88.57) \\
\hline
\end{tabular}

\end{table}

\begin{figure}[htbp]
\centering
\includegraphics[width=0.49\textwidth]{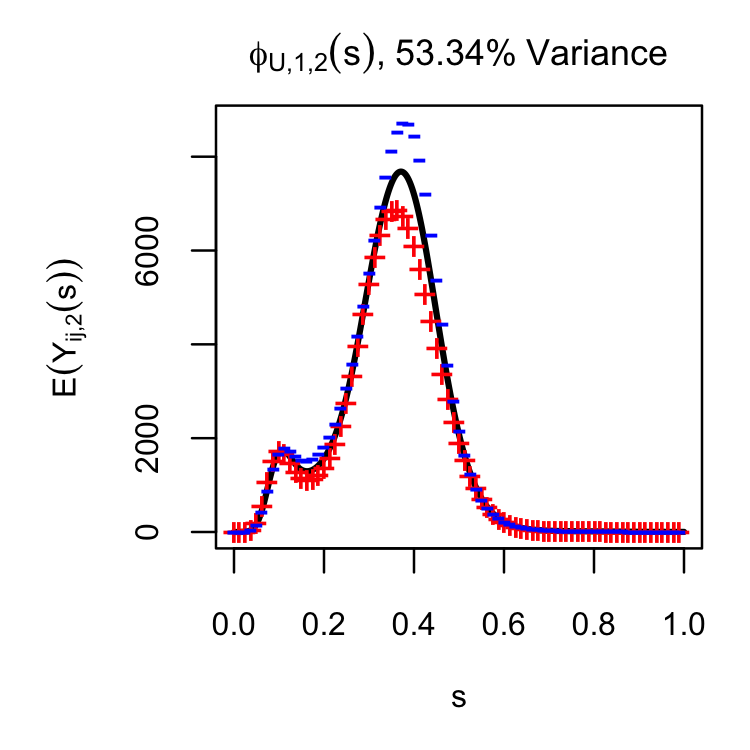}
 \includegraphics[width=0.49\textwidth]{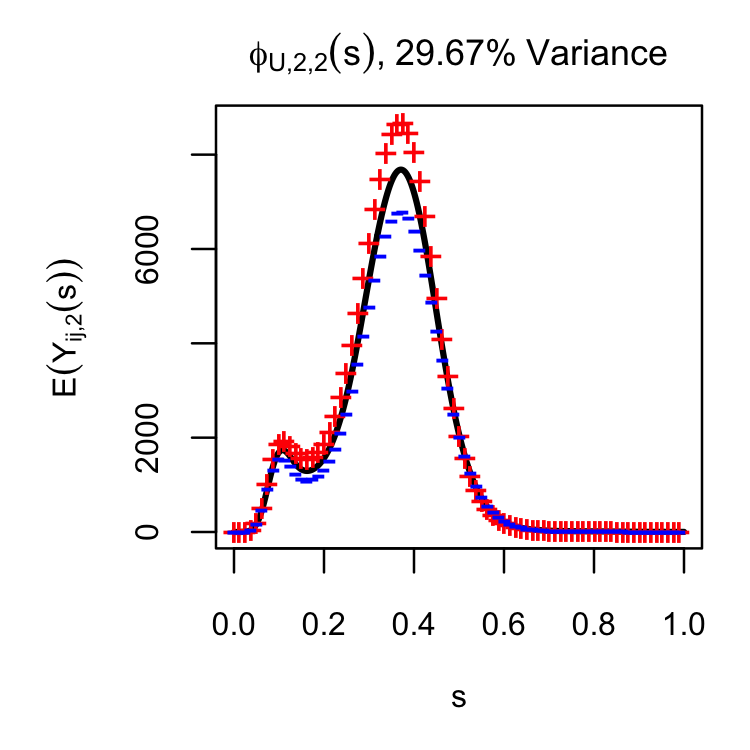}
\hfill
\includegraphics[width=0.49\textwidth]{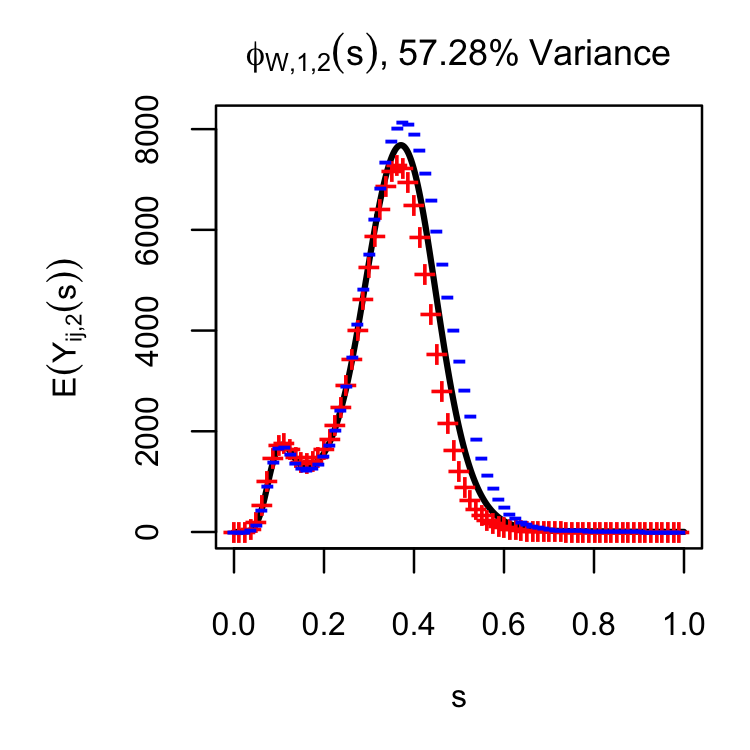}
\includegraphics[width=0.49\textwidth]{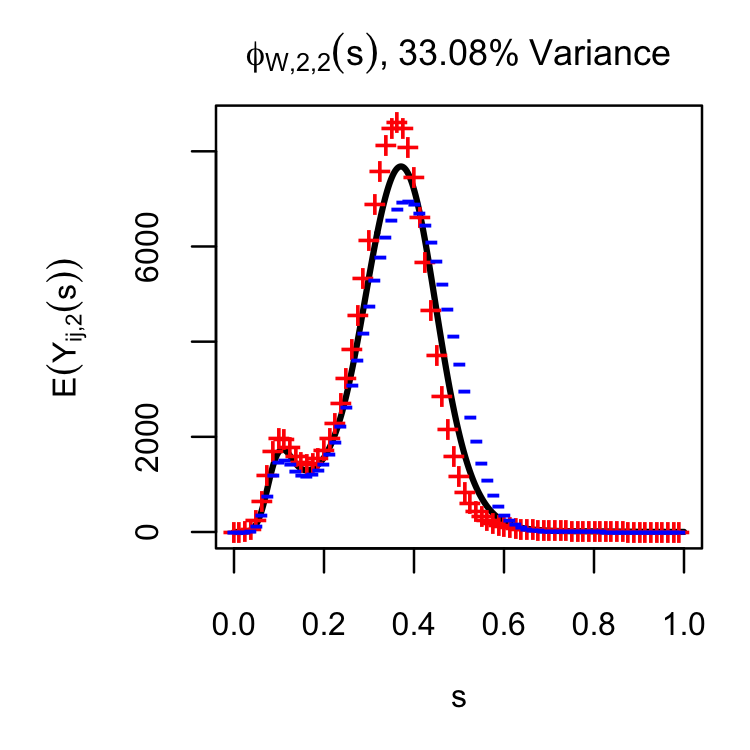}
\caption{Response-scale representations of the first two subject-level and visit-level functional principal components for $H_2$. The top row displays the first two subject-level components, and the bottom row displays the first two visit-level components. The black curve represents the reference expected Betti curve, while the red \(+\) and blue \(-\) symbols represent positive and negative deviations along each functional principal component direction.} 
    \label{fig:fpc_h2}
\end{figure}

After assessing the relative contributions of subject- and visit-level variability, we next examine the dominant patterns of variation within each level using functional principal components. 
To facilitate interpretation of the modes of variation on the original count scale, Figure~\ref{fig:fpc_h2} illustrates the first two subject- and visit-level components for $H_2$. For subject-level component $k$, the corresponding curves are represented as
$\exp\left\{
\beta_{0,2}(s)
\pm
\sqrt{\lambda_{U,k,2}}\,
\phi_{U,k,2}(s)
\right\},$
and for the visit-level component $\ell$,they are given by
$\exp\left\{
\beta_{0,2}(s)
\pm
\sqrt{\lambda_{W,\ell,2}}\,
\phi_{W,\ell,2}(s)
\right\}.$ The positive and negative directions represent one-standard deviation along each component and illustrate how variation along each component modifies the reference expected $H_2$ Betti curve. \\ 









The first subject-level component explains 53.34\% of the variation and primarily reflects differences in the prominence of the dominant peak, while the second component explains 29.67\% and captures additional changes in the shape and relative prominence of the early and dominant peak regions. A similar pattern is observed at the visit level, where the first two components explain 57.28\% and 33.08\% of the variation and are associated with pronounced changes in the dominant peak, including its shape and position over $s$. The corresponding $H_0$ and $H_1$ component plots are provided in Figure~16 and~17 of the Appendix, respectively. Compared with $H_0$ and $H_1$, the $H_2$ variation is less concentrated in the first component, indicating that several
distinct modes contribute to both between-subject and within-subject topological variation. 


\section{Discussion}
\label{sec:6}

In this study, we present a generalized multilevel functional framework for analyzing longitudinal topological summaries derived from three-dimensional structural MRI. Rather than representing persistent-homology information using a single summary measure, the proposed framework treats Betti curves as count-valued longitudinal functional outcomes and models their variation across the filtration continuum. This formulation provides a direct statistical approach for investigating how topological features of brain structure vary with demographic and clinical factors while accounting for repeated observations within subjects. The negative-binomial model accommodates the discrete and overdispersed nature of Betti counts, while the multilevel functional representation separates variation between subjects from additional variation across visits. The Bayesian implementation further allows joint estimation of the functional regression effects and multilevel functional principal components, while providing uncertainty quantification for both population-level covariate effects and latent modes of variation.\\

The application to longitudinal MRI data from the OASIS-2 study illustrates how the proposed framework can be used to examine associations between topological features and subject-level and clinical characteristics while preserving their variation across the filtration continuum. The functional regression results show that Age, gender, follow-up time, and dementia severity are associated with changes in the log expected Betti counts at specific regions of the filtration domain, with the magnitude, location, and shape of these changes differing across homology dimensions. 
An important contribution of the multilevel formulation is its ability to quantify the relative contributions of subject- and visit-level functional variability. In the OASIS-2 application, visit-level variation accounts for the majority of the unexplained functional variability for all three homology dimensions. However, notice that the contribution of subject-level variation increases for higher homology dimensions. Thus, although variation across visits within subjects remains dominant, between-subject differences become increasingly prominent in the higher-dimensional topological features. The functional principal component analysis provides a complementary description of these sources of variation by identifying dominant patterns of change in the shape and magnitude of the Betti curves. The first few components explain a substantial proportion of the variation at both the subject and visit levels, while the higher-dimensional $H_2$ curves exhibit a less concentrated variation structure than $H_0$ and $H_1$.\\

Several avenues remain for future research. First, the current analysis considers $H_0$, $H_1$, and $H_2$ separately; joint modeling of multiple homology dimensions could account for their dependence and provide a more comprehensive characterization of disease-related morphological changes. 
Second, the current analysis focuses on structural MRI alone. Extending the proposed approach to multimodal longitudinal imaging, including diffusion and functional MRI, could enable joint investigation of topological changes across complementary aspects of brain organization.

\bibliographystyle{unsrtnat}
\bibliography{LFDA}  

\section{Appendix}
\section*{A.1 Demographic and Clinical Characteristics of the Participants}

Table~\ref{tab:S1_characteristics} summarizes the baseline demographic
and clinical characteristics of the participants included in the analysis.

\begin{table}[ht]
\centering
\caption{Demographic and clinical characteristics of the analysis sample. CDR denotes the Clinical Dementia Rating.}
\label{tab:S1_characteristics}
\begin{tabular}{lc}
\hline
Characteristic & Summary \\
\hline
\multicolumn{2}{l}{\textit{Demographic characteristics}} \\
\quad Participants, $n$              & 22 \\
\quad Female, $n$ (\%)                & 8 (36.4) \\
\quad Age at baseline, years (standard deviation)         & 76.1 (6.6) \\[2pt]

\multicolumn{2}{l}{\textit{Baseline clinical status}} \\
\quad CDR $=0$, number of observation                  & 7 \\
\quad CDR $=0.5$, number of observation                 & 12 \\
\quad CDR $=1$, number of observation                   & 3 \\[2pt]

\multicolumn{2}{l}{\textit{Follow-up}} \\
\quad Follow-up duration, years (standard deviation)       & 3.9 (1.2) \\
\hline
\end{tabular}
\footnotesize
\end{table}

\newpage
\section*{A.2 Observed Betti curves across second and third visits}

\begin{figure}[htbp]
    \centering

    \begin{minipage}[t]{0.32\textwidth}
        \centering
        \includegraphics[width=\linewidth]{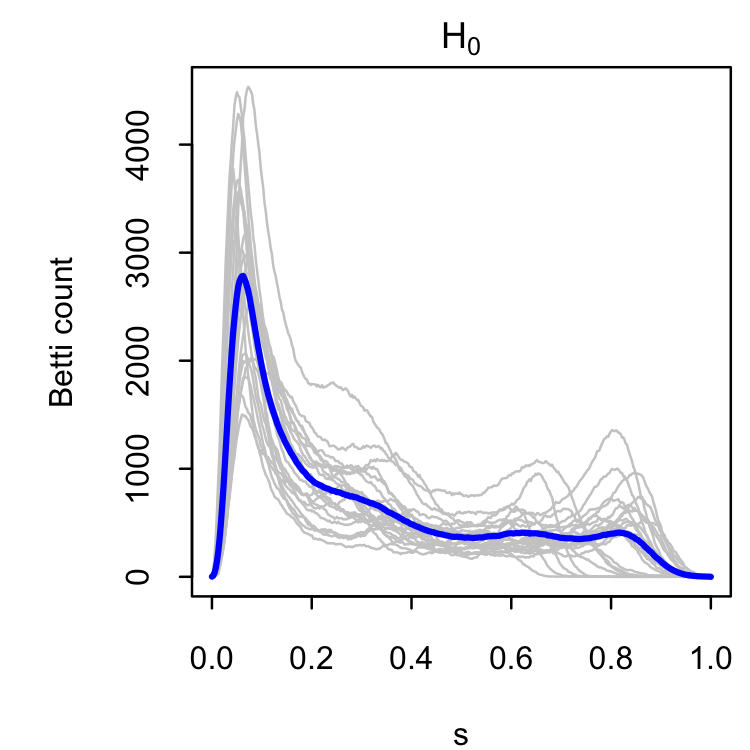}
    \end{minipage}
    \hspace{-0.0025\textwidth}
    \begin{minipage}[t]{0.32\textwidth}
        \centering
        \includegraphics[width=\linewidth]{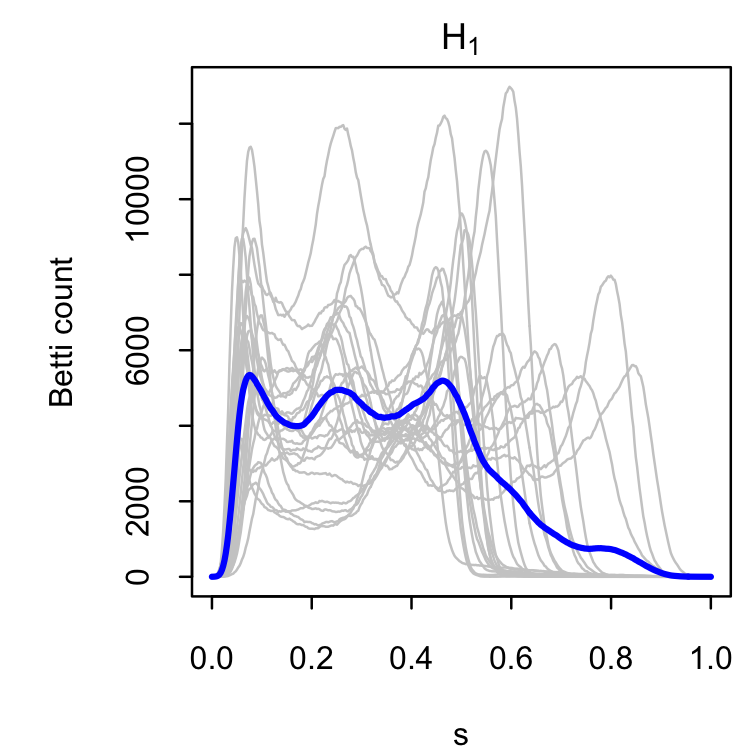}
    \end{minipage}
    \hspace{-0.0025\textwidth}
    \begin{minipage}[t]{0.32\textwidth}
        \centering
        \includegraphics[width=\linewidth]{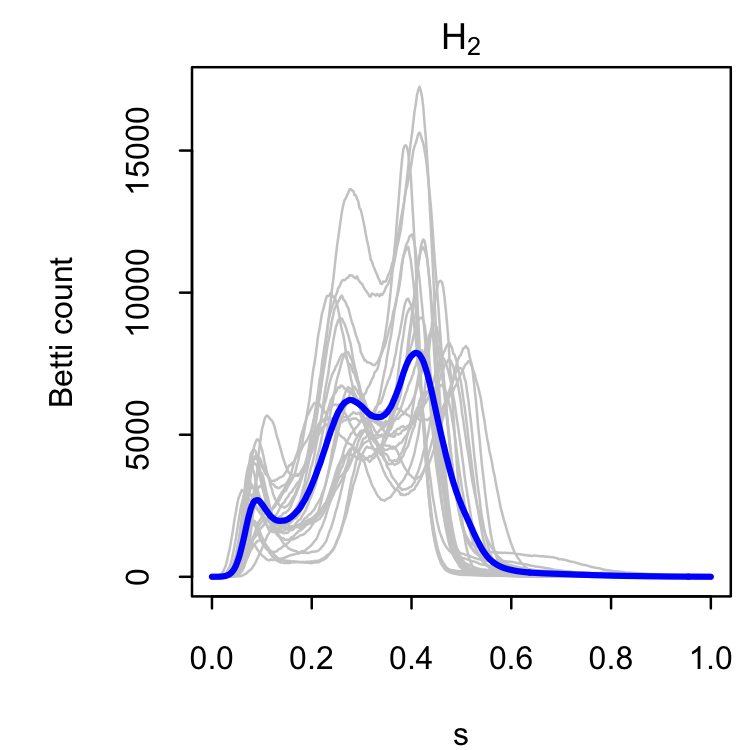}
    \end{minipage}

    \caption{Observed $H_0$, $H_1$, $H_2$ Betti curves at the second visit. Gray lines represent individual subject curves, and the blue line represents the mean Betti curve across subjects.}
    \label{fig:visit_1}
\end{figure}

The second-visit Betti curves exhibit overall patterns of between-subject variation similar to those observed at the first visit, with \(H_1\) showing the most pronounced variability across subjects.


\begin{figure}[htbp]
    \centering

    \begin{minipage}[t]{0.32\textwidth}
        \centering
        \includegraphics[width=\linewidth]{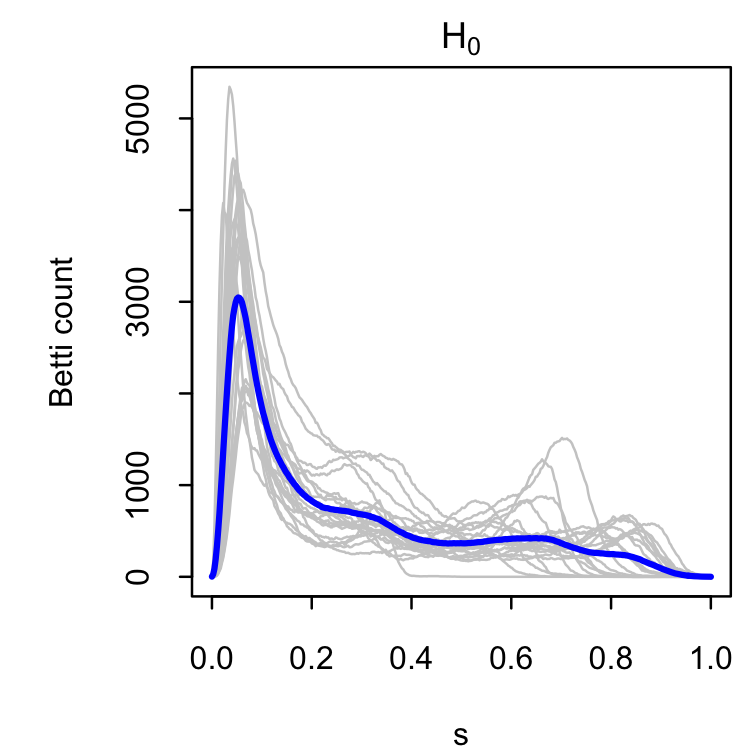}
    \end{minipage}
    \hspace{-0.0025\textwidth}
    \begin{minipage}[t]{0.32\textwidth}
        \centering
        \includegraphics[width=\linewidth]{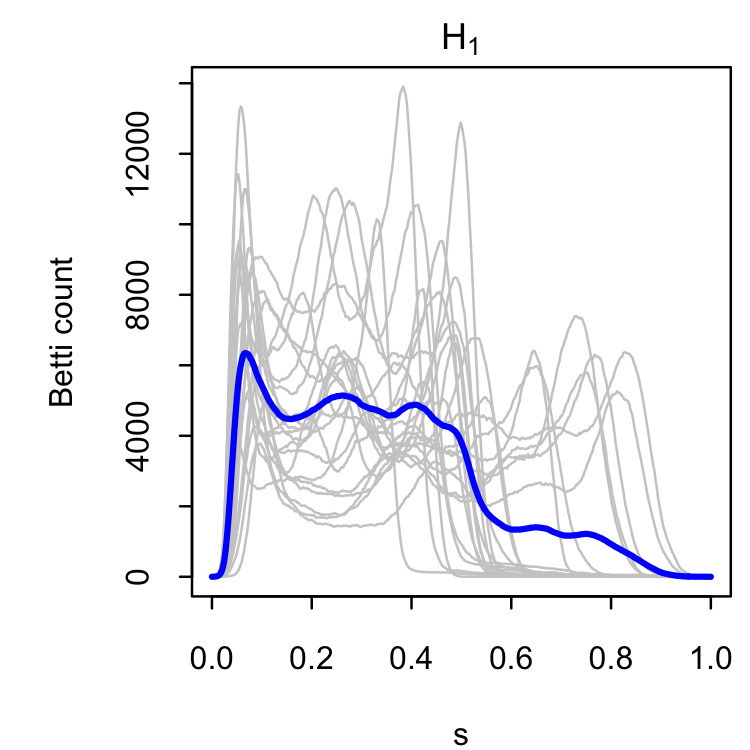}
    \end{minipage}
    \hspace{-0.0025\textwidth}
    \begin{minipage}[t]{0.32\textwidth}
        \centering
        \includegraphics[width=\linewidth]{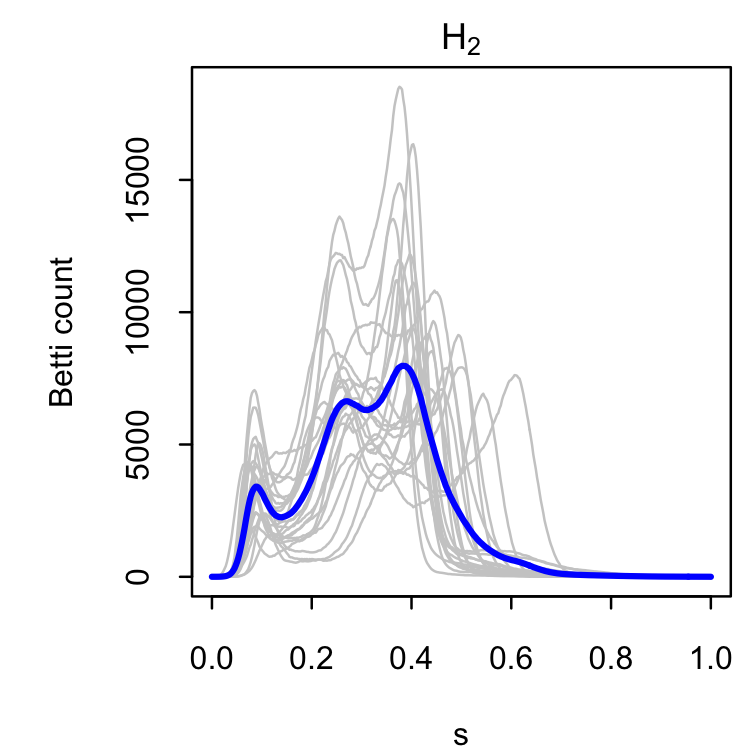}
    \end{minipage}

    \caption{Observed $H_0$, $H_1$, $H_2$ Betti curves at the third visit. Gray lines represent individual subject curves, and the blue line represents the mean Betti curve across subjects.}
    \label{fig:visit_1}
\end{figure}

The third-visit Betti curves show comparable overall patterns, with substantial between-subject variability, particularly for \(H_1\).
\clearpage
\newpage

\section*{A.3 Fixed effect coefficient function for $H_0$}

\begin{figure}[htbp]
    \centering
    \begin{minipage}[t]{0.33\textwidth}
        \centering
        \includegraphics[width=\linewidth]{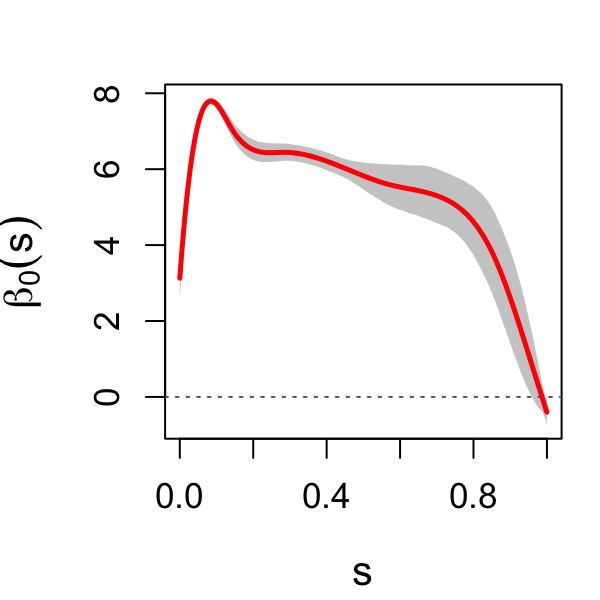}
    \end{minipage}
    \hfill
    \begin{minipage}[t]{0.33\textwidth}
        \centering
        \includegraphics[width=\linewidth]{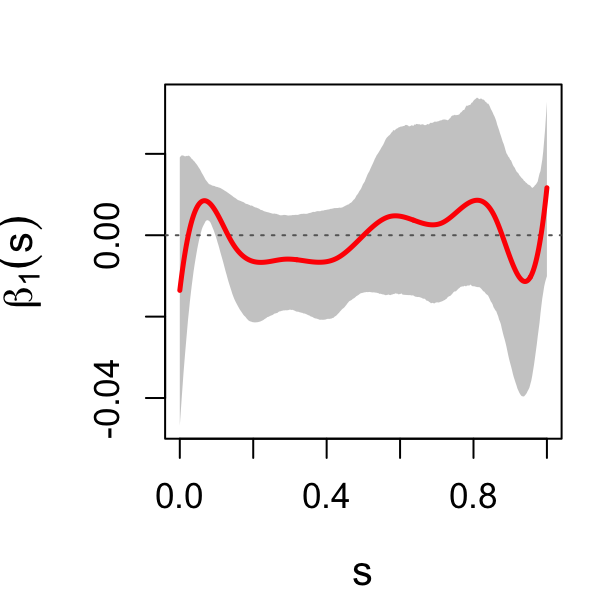}
    \end{minipage}
    \hfill
    \begin{minipage}[t]{0.33\textwidth}
        \centering
        \includegraphics[width=\linewidth]{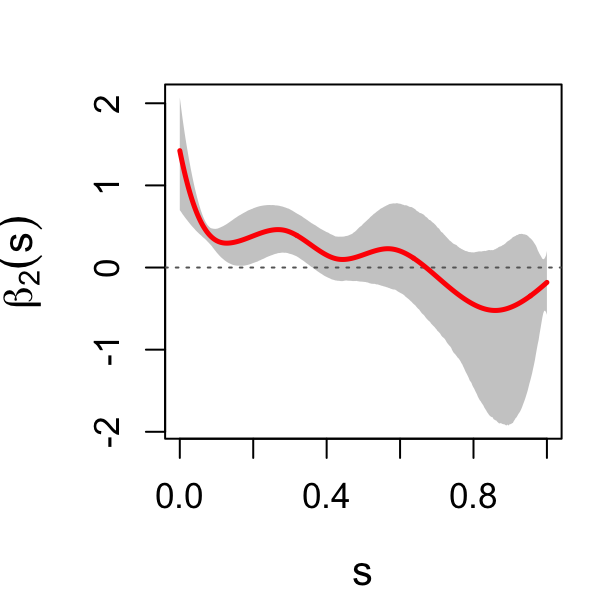}
    \end{minipage}
    \hfill
    \begin{minipage}[t]{0.33\textwidth}
        \centering
        \includegraphics[width=\linewidth]{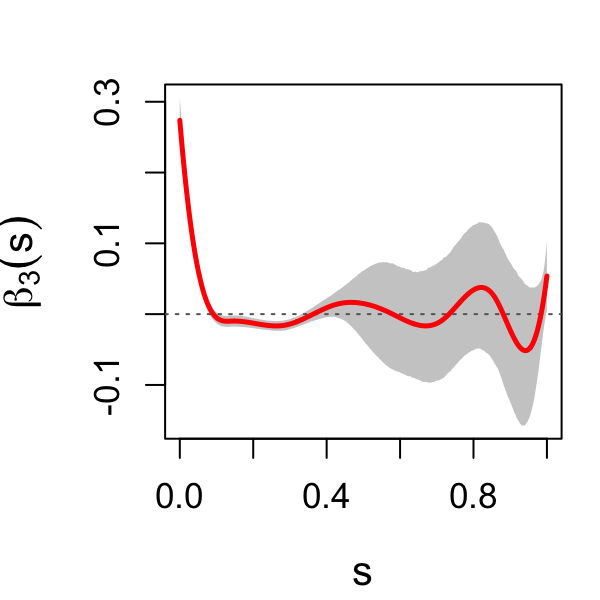}
    \end{minipage}
    \hfill
    \begin{minipage}[t]{0.33\textwidth}
        \centering
        \includegraphics[width=\linewidth]{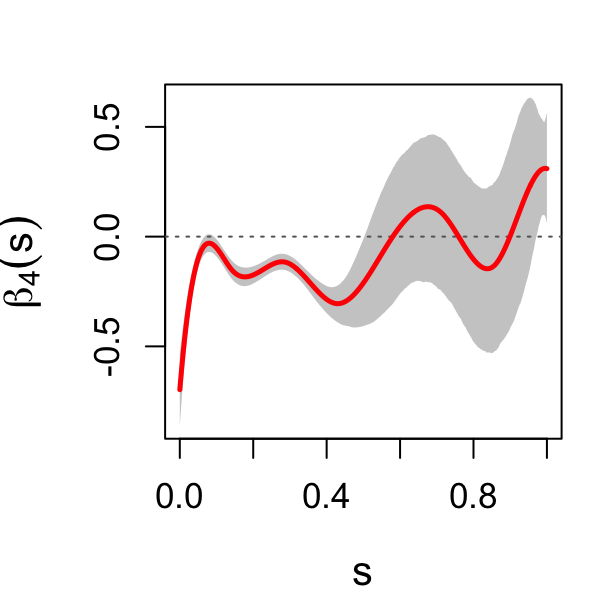}
    \end{minipage}
    \hfill
    \begin{minipage}[t]{0.33\textwidth}
        \centering
        \includegraphics[width=\linewidth]{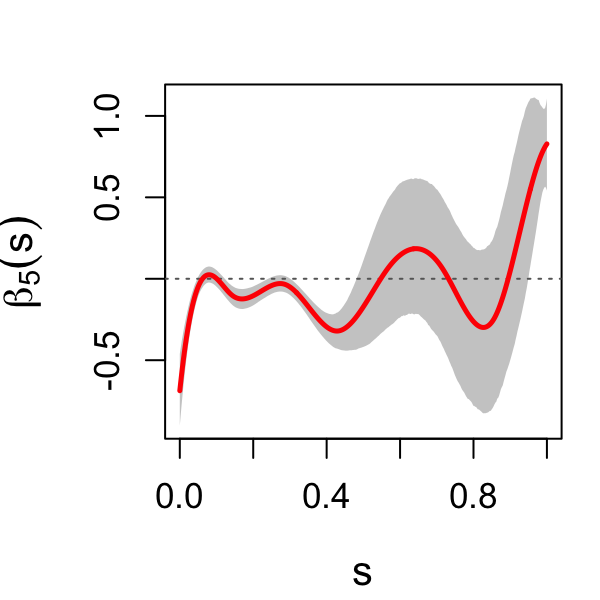}
    \end{minipage}
    \caption{Posterior mean coefficient functions and corresponding pointwise 95\% credible intervals for $H_0$.}
    \label{fig:coef_h2}
\end{figure}

For $H_0$ Betti curves, the strongest coefficient patterns are observed for gender and follow-up time at lower filtration values, whereas the baseline-age and CDR coefficients remain closer to zero over much of the filtration domain.

\newpage
\section*{A.4 Posterior convergence diagnostics}

Posterior convergence is assessed using the potential scale reduction statistic $\widehat{R}$ and the effective sample size (ESS). Table~\ref{tab:convergence} summarizes the maximum $\widehat{R}$ and minimum effective sample size across the model parameters for each homology dimension. No divergent transitions or maximum-treedepth exceedances are observed for any of the fitted models.

\begin{table}[htbp]
\centering
\caption{Posterior convergence diagnostics for the generalized multilevel
functional models fitted to the $H_0$, $H_1$, and $H_2$ Betti curves.}
\label{tab:convergence}
\begin{tabular}{lcc}
\hline
\textbf{Homology dimension} & \textbf{Maximum $\widehat{R}$} & \textbf{Minimum ESS} \\
\hline
$H_0$ & 1.005 & 904  \\
$H_1$ & 1.002 & 1116 \\
$H_2$ & 1.005 & 1038 \\
\hline
\end{tabular}
\end{table}

\begin{figure}[htbp]
    \centering
    \begin{minipage}[t]{0.49\textwidth}
        \centering
        \includegraphics[width=\linewidth]{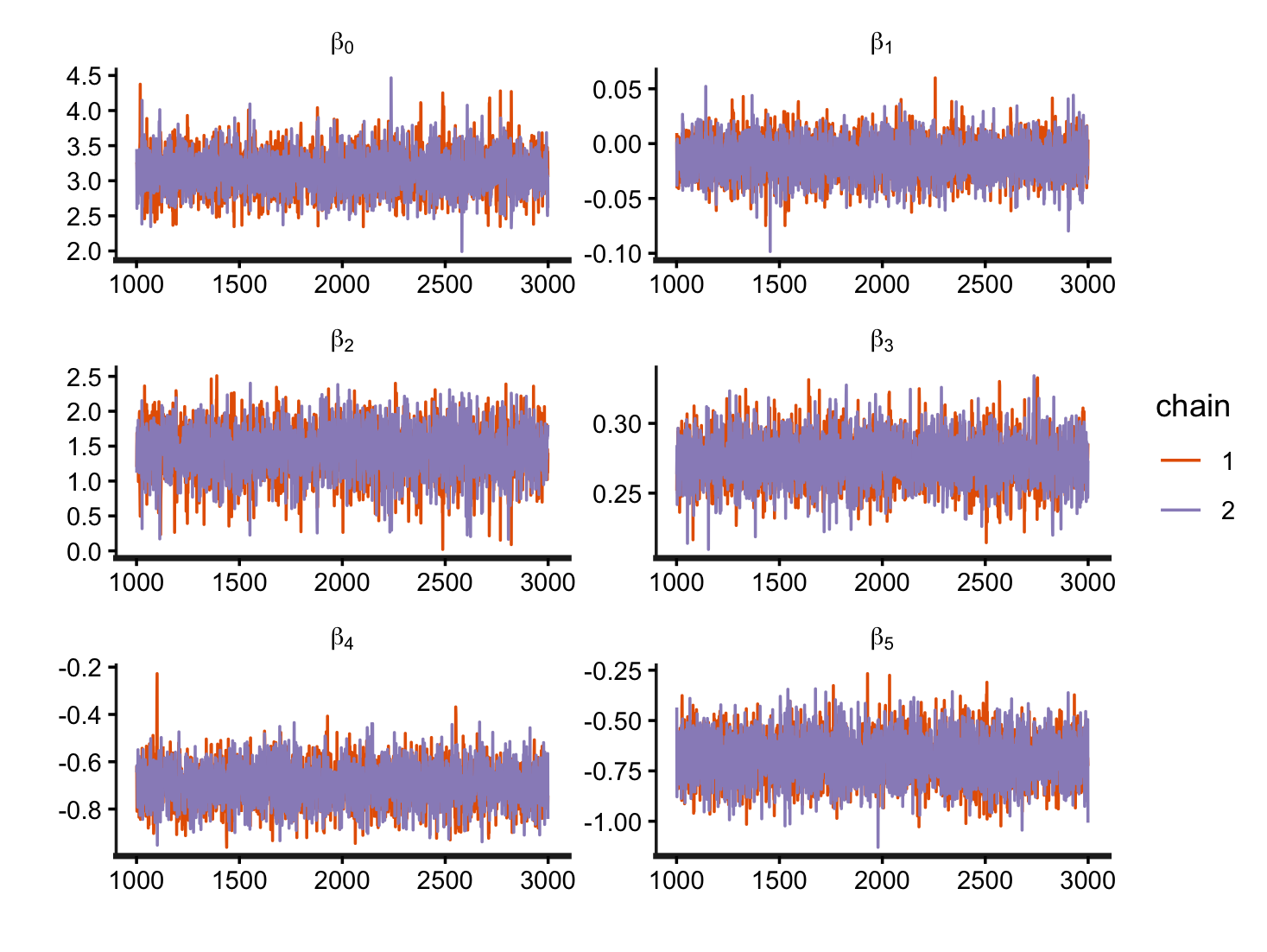}
    \end{minipage}
    \hfill
    \begin{minipage}[t]{0.49\textwidth}
        \centering
        \includegraphics[width=\linewidth]{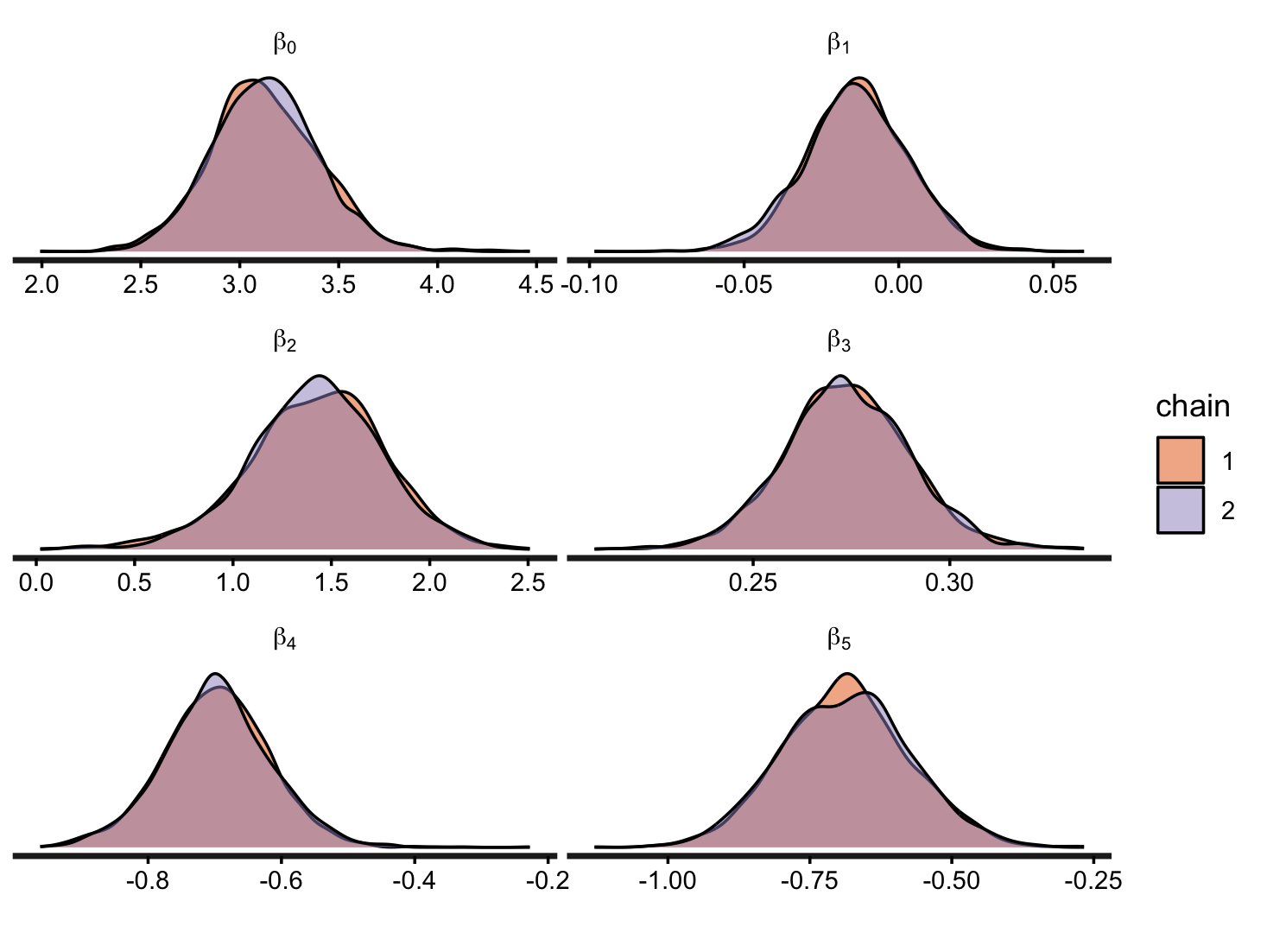}
    \end{minipage}
    \hfill
    \begin{minipage}[t]{0.49\textwidth}
        \centering
        \includegraphics[width=\linewidth]{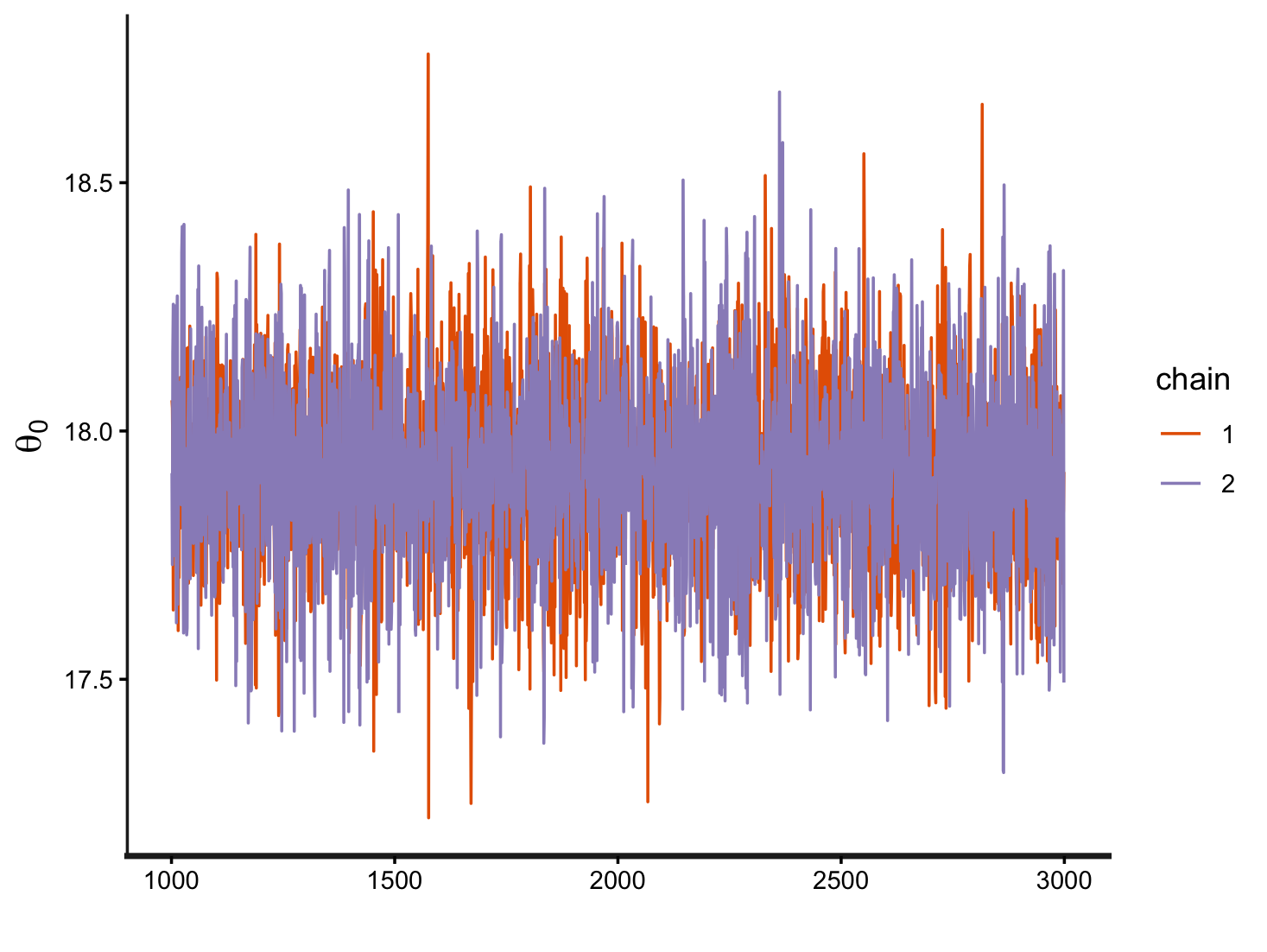}
    \end{minipage}
    \hfill
    \begin{minipage}[t]{0.49\textwidth}
        \centering
        \includegraphics[width=\linewidth]{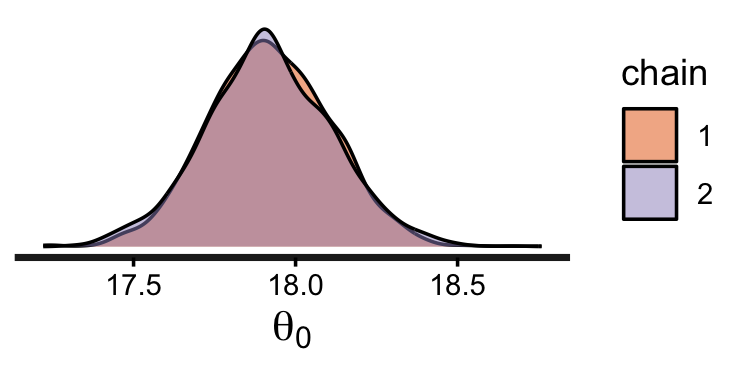}
    \end{minipage}
    \caption{Representative MCMC trace plots and chain-specific posterior density estimates for selected fixed-effect coefficients and the negative-binomial dispersion parameter $\theta_0$ for $H_0$.}
    \label{fig:diag_h0}
\end{figure}

\begin{figure}[htbp]
    \centering
    \begin{minipage}[t]{0.49\textwidth}
        \centering
        \includegraphics[width=\linewidth]{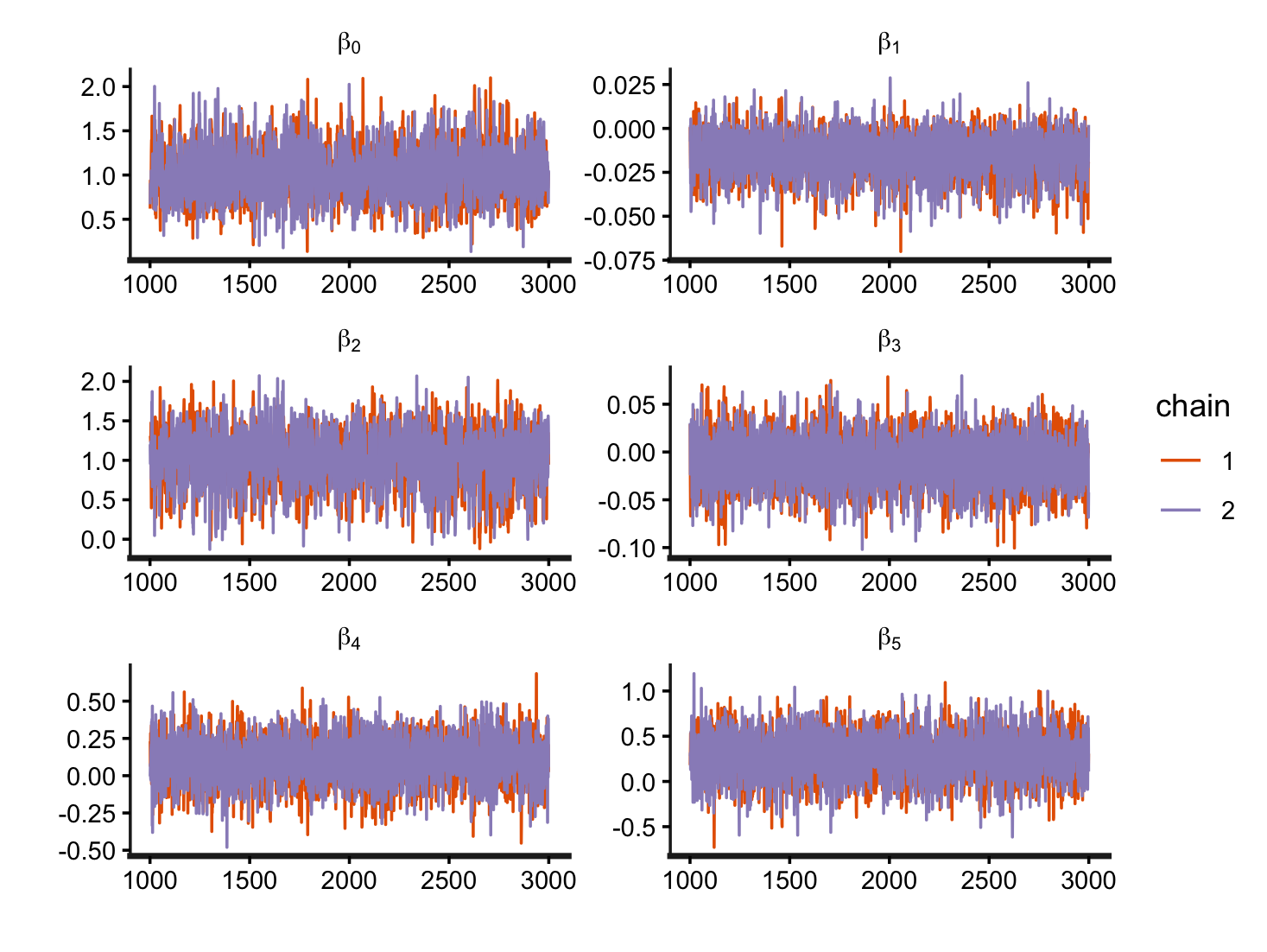}
    \end{minipage}
    \hfill
    \begin{minipage}[t]{0.49\textwidth}
        \centering
        \includegraphics[width=\linewidth]{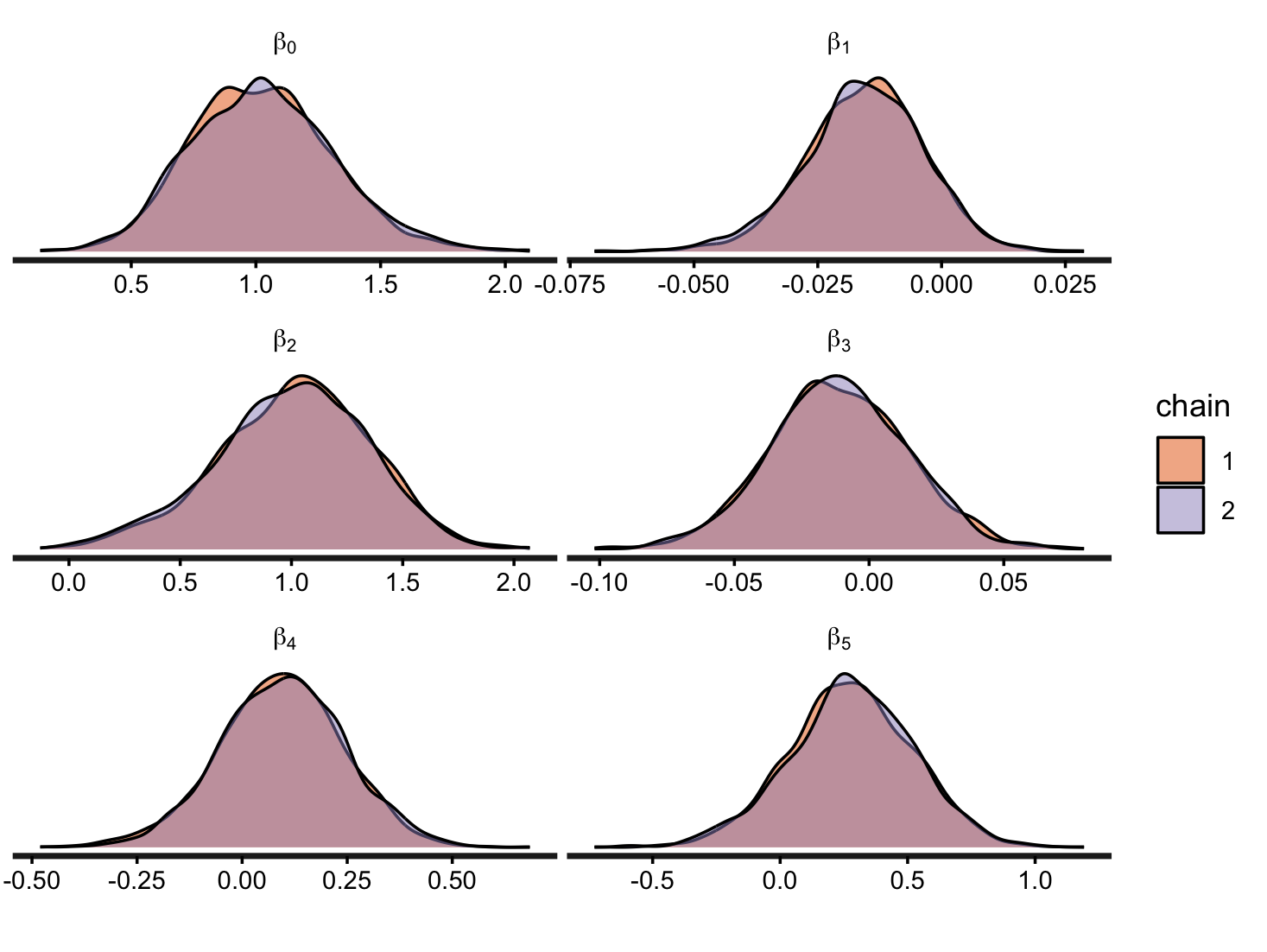}
    \end{minipage}
    \hfill
    \begin{minipage}[t]{0.49\textwidth}
        \centering
        \includegraphics[width=\linewidth]{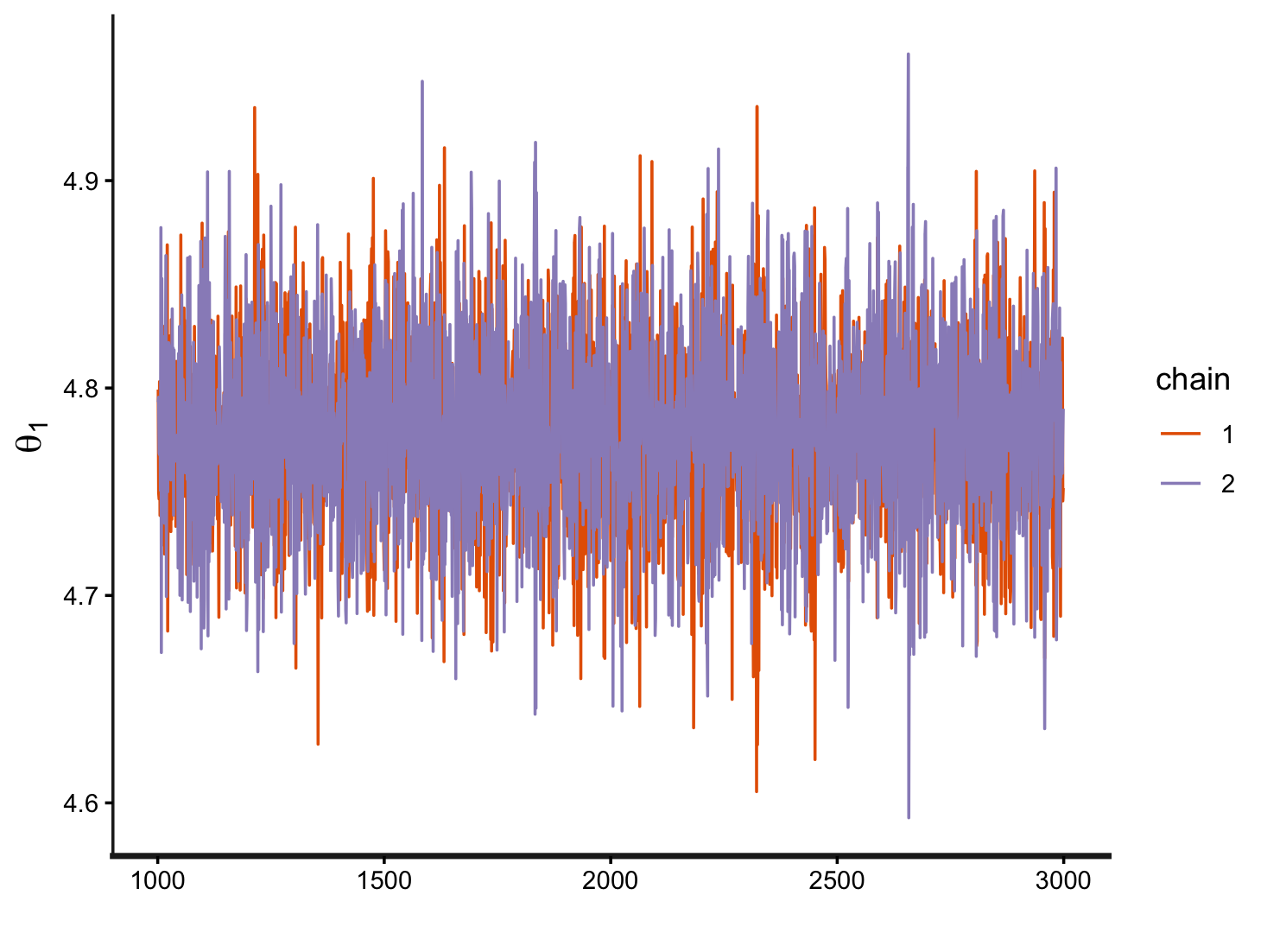}
    \end{minipage}
    \hfill
    \begin{minipage}[t]{0.49\textwidth}
        \centering
        \includegraphics[width=\linewidth]{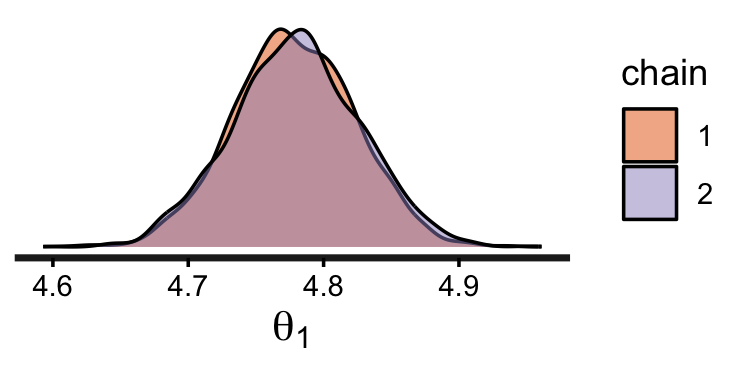}
    \end{minipage}
    \caption{Representative MCMC trace plots and chain-specific posterior density estimates for selected fixed-effect coefficients and the negative-binomial dispersion parameter $\theta_1$ for $H_1$.}
    \label{fig:diag_h1}
\end{figure}

\begin{figure}[htbp]
    \centering
    \begin{minipage}[t]{0.49\textwidth}
        \centering
        \includegraphics[width=\linewidth]{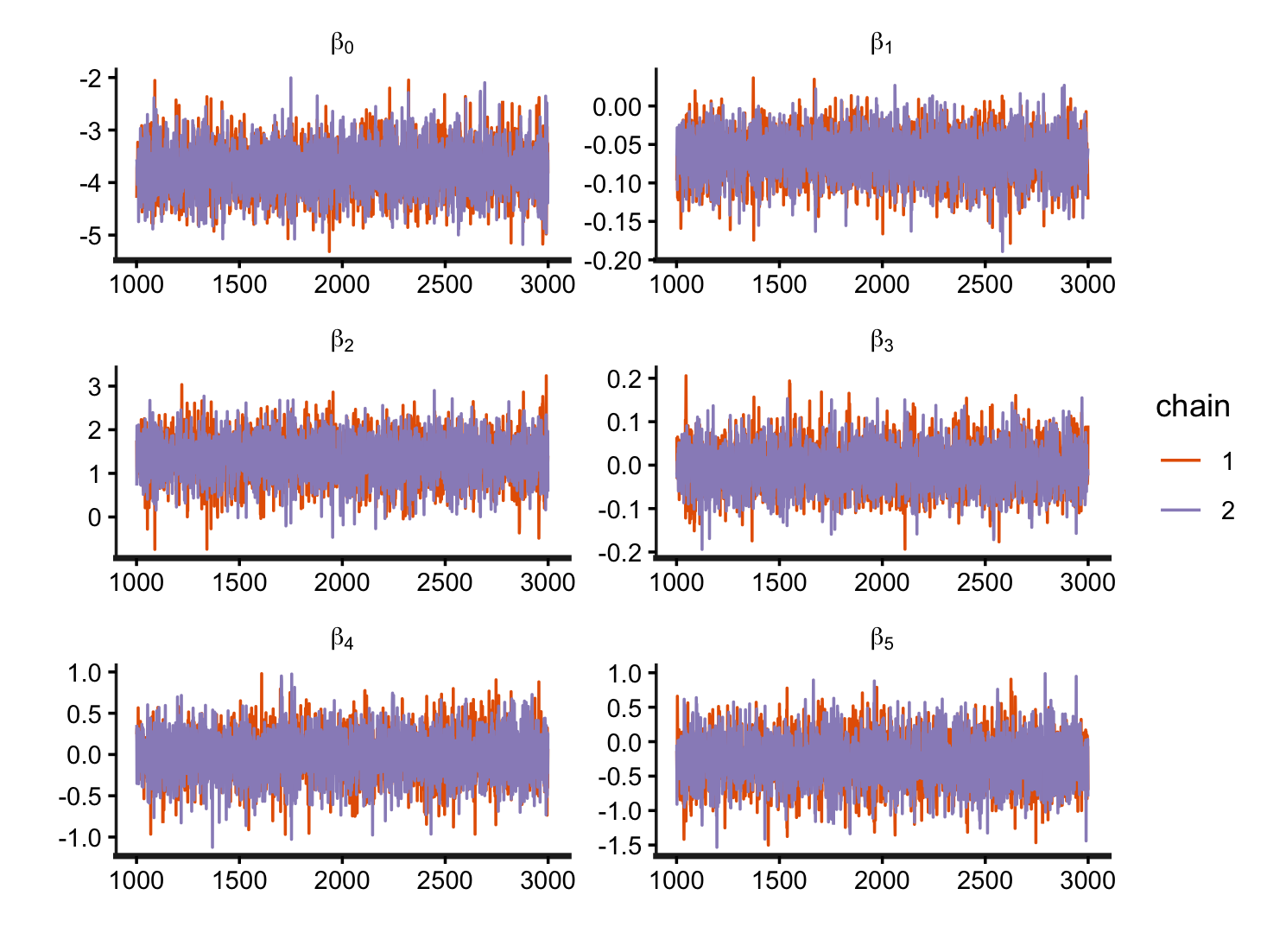}
    \end{minipage}
    \hfill
    \begin{minipage}[t]{0.49\textwidth}
        \centering
        \includegraphics[width=\linewidth]{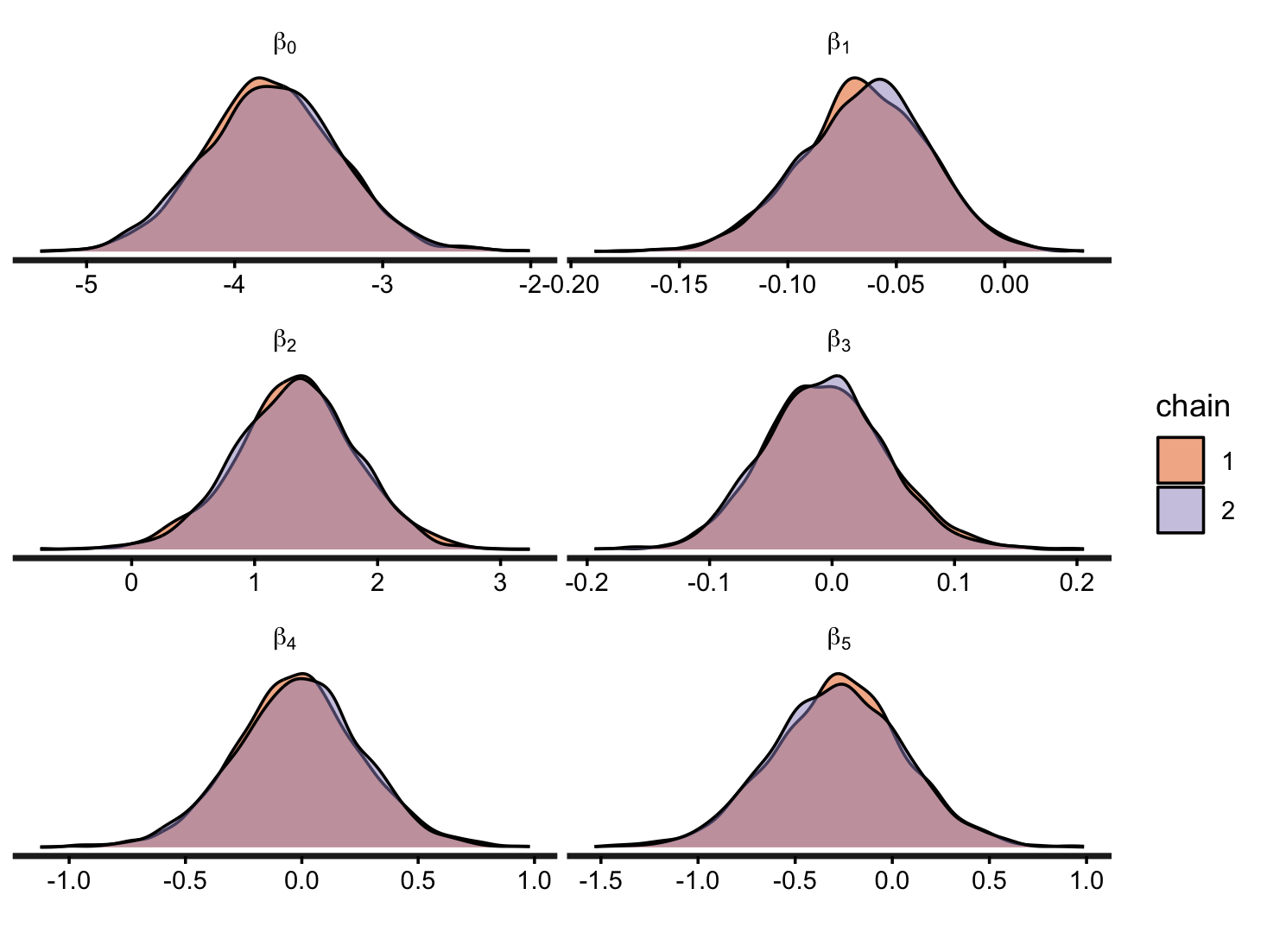}
    \end{minipage}
    \hfill
    \begin{minipage}[t]{0.49\textwidth}
        \centering
        \includegraphics[width=\linewidth]{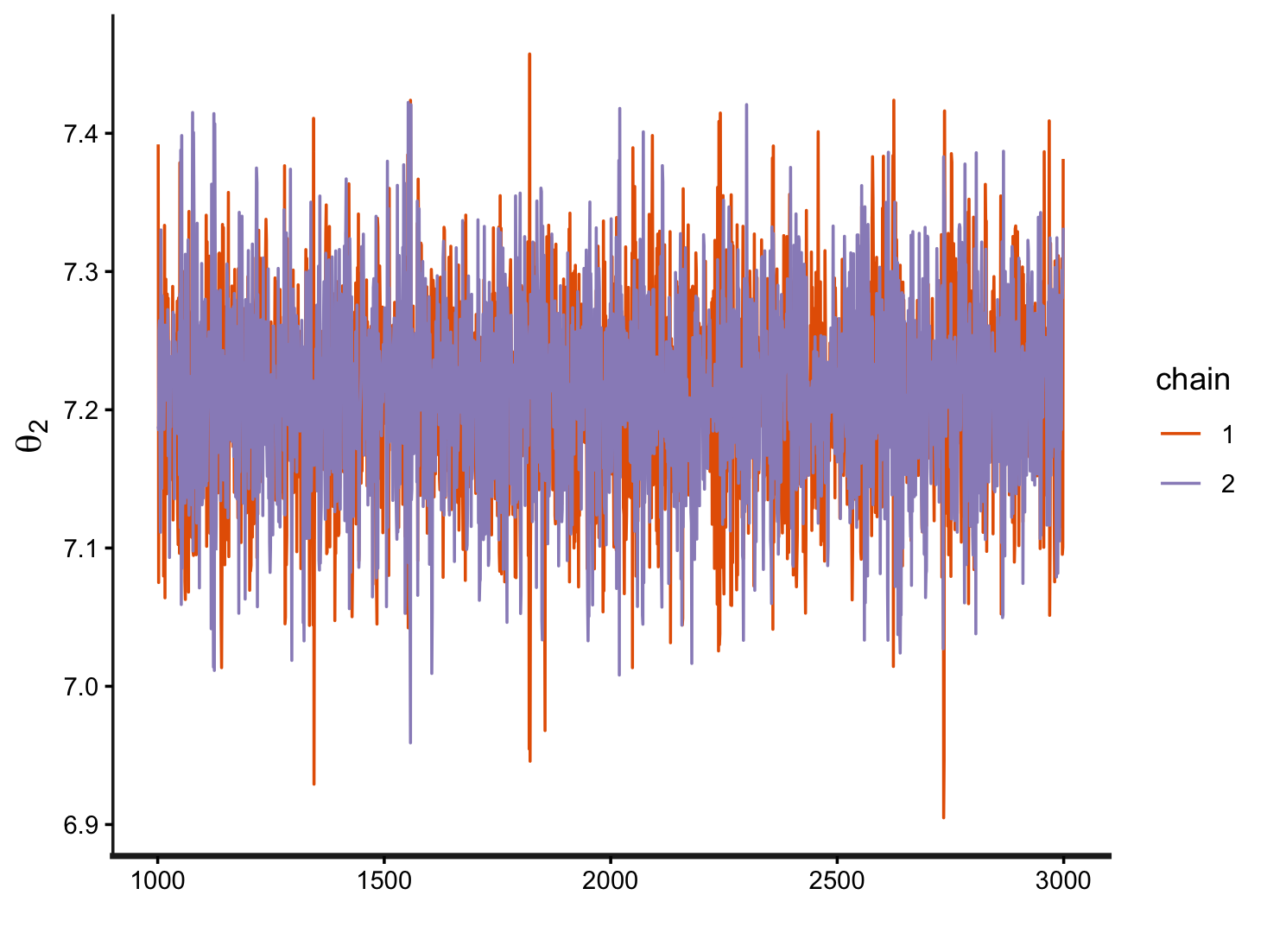}
    \end{minipage}
    \hfill
    \begin{minipage}[t]{0.49\textwidth}
        \centering
        \includegraphics[width=\linewidth]{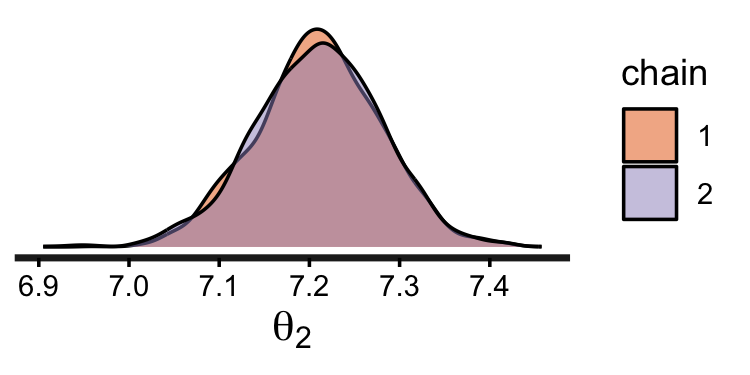}
    \end{minipage}
    \caption{Representative MCMC trace plots and chain-specific posterior density estimates for selected fixed-effect coefficients and the negative-binomial dispersion parameter $\theta_2$ for $H_2$.}
    \label{fig:diag_h2}
\end{figure}

Figures~\ref{fig:diag_h0}--~\ref{fig:diag_h2} present representative trace plots and chain-specific posterior density estimates for $H_0$, $H_1$, and $H_2$, respectively. The trace plots show stable mixing between chains, while the corresponding posterior density estimates exhibit substantial overlap.

\clearpage
\newpage
\section*{A.5 Response-scale covariate associations for $H_0$}
\label{sec:supp_h0_response}

\begin{figure}[htbp]
    \centering
    \includegraphics[width=0.8\linewidth]{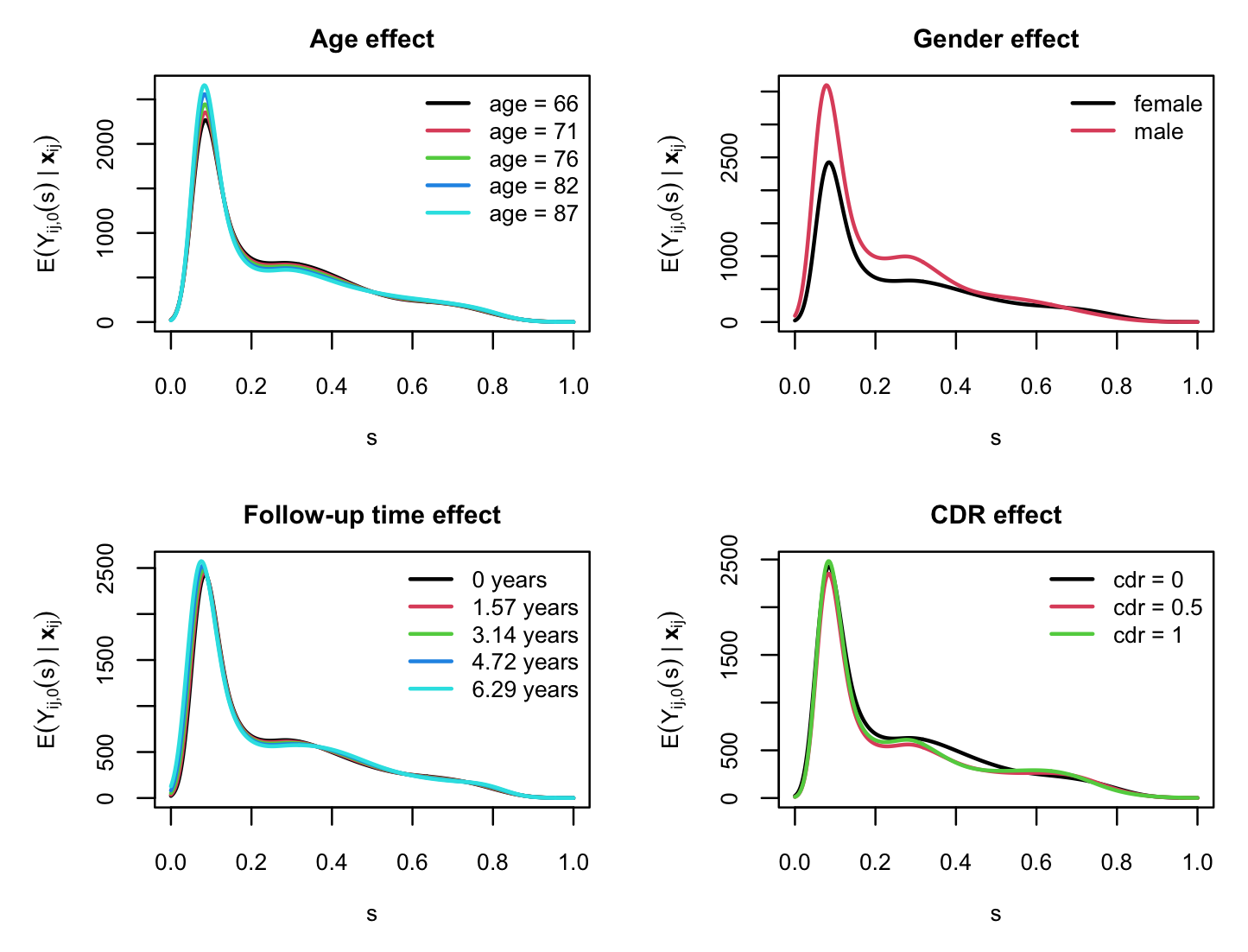}
    \caption{Estimated population-level $H_0$ Betti curves on the original count scale for selected values of baseline age, gender, follow-up time, and CDR. Within each panel, the remaining covariates are held at their reference values, and the subject- and visit-level latent effects are set to zero.}
    \label{fig:coef_h0_effect}
\end{figure}

For \(H_0\), the response-scale curves show a clear gender difference near the dominant early peak, whereas the curves remain unchanged across baseline age, follow-up time, and CDR.

\clearpage
\newpage
\section*{A.6 Functional principal component results for $H_0$ and $H_1$}

\begin{figure}[htbp]
    \centering
    \begin{minipage}[t]{0.49\textwidth}
        \centering
        \includegraphics[width=\linewidth]{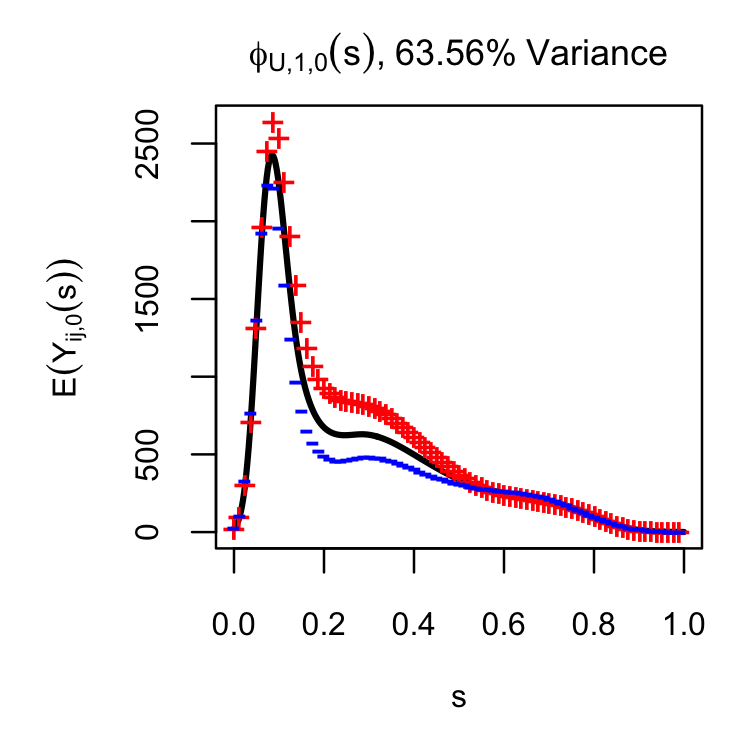}
    \end{minipage}
    \hfill
    \begin{minipage}[t]{0.49\textwidth}
        \centering
        \includegraphics[width=\linewidth]{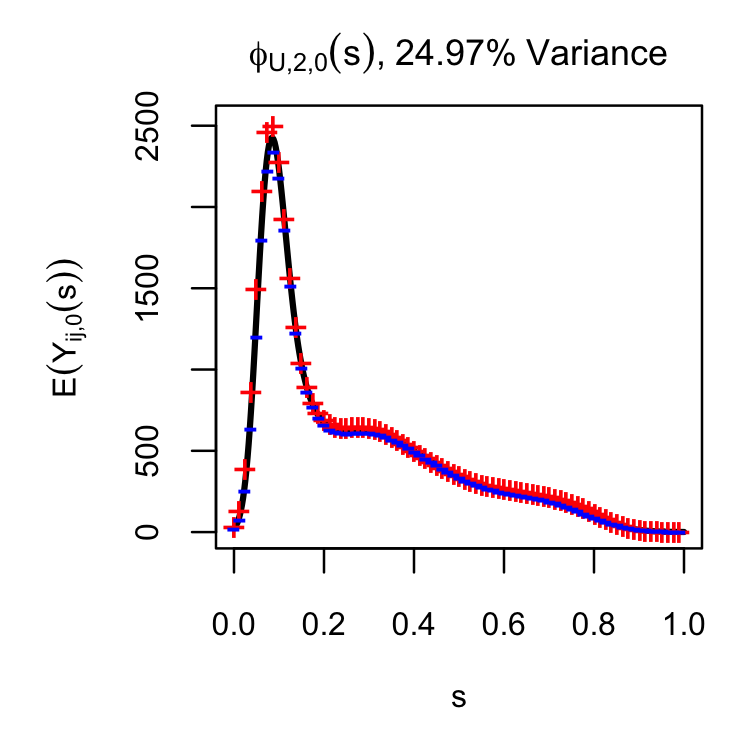}
    \end{minipage}
    \hfill
    \begin{minipage}[t]{0.49\textwidth}
        \centering
        \includegraphics[width=\linewidth]{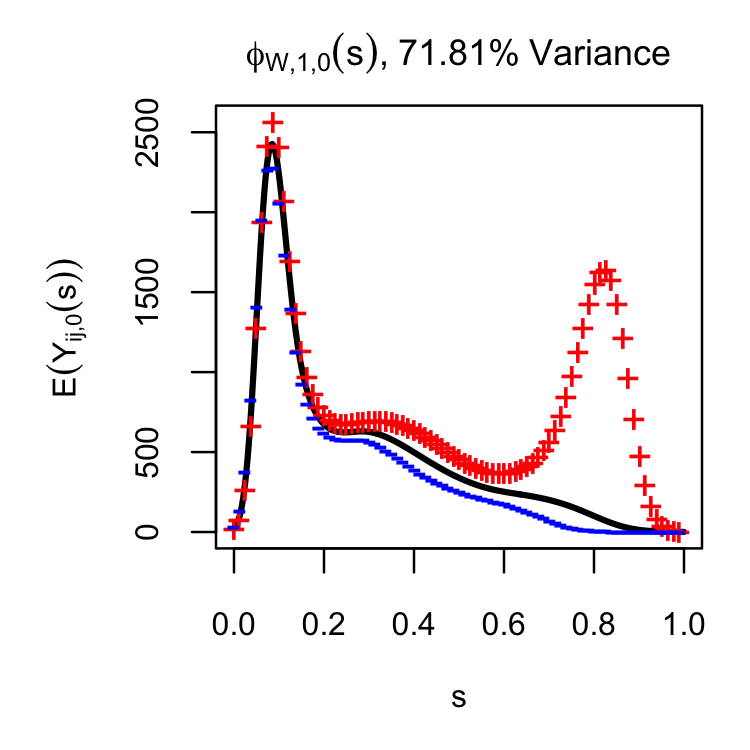}
    \end{minipage}
    \hfill
    \begin{minipage}[t]{0.49\textwidth}
        \centering
        \includegraphics[width=\linewidth]{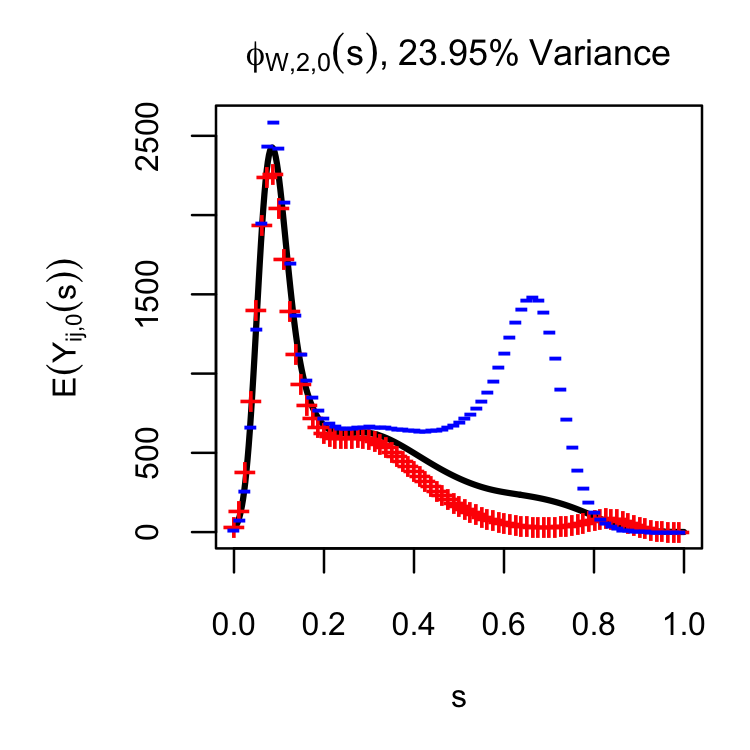}
    \end{minipage}
    \caption{Response-scale representations of the first two subject-level and visit-level functional principal components for $H_0$. The top row displays the first two subject-level components, and the bottom row displays the first two visit-level components. The black curve represents the reference expected Betti curve, while the red \(+\) and blue \(-\) symbols represent positive and negative deviations along each functional principal component direction.}
    \label{fig:fpc_h0}
\end{figure}

For \(H_0\), the leading subject-level components primarily reflect variation in the magnitude of the early peak, whereas the visit-level components capture additional changes across the middle and higher filtration values.

\newpage
\begin{figure}[htbp]
    \centering
    \begin{minipage}[t]{0.49\textwidth}
        \centering
        \includegraphics[width=\linewidth]{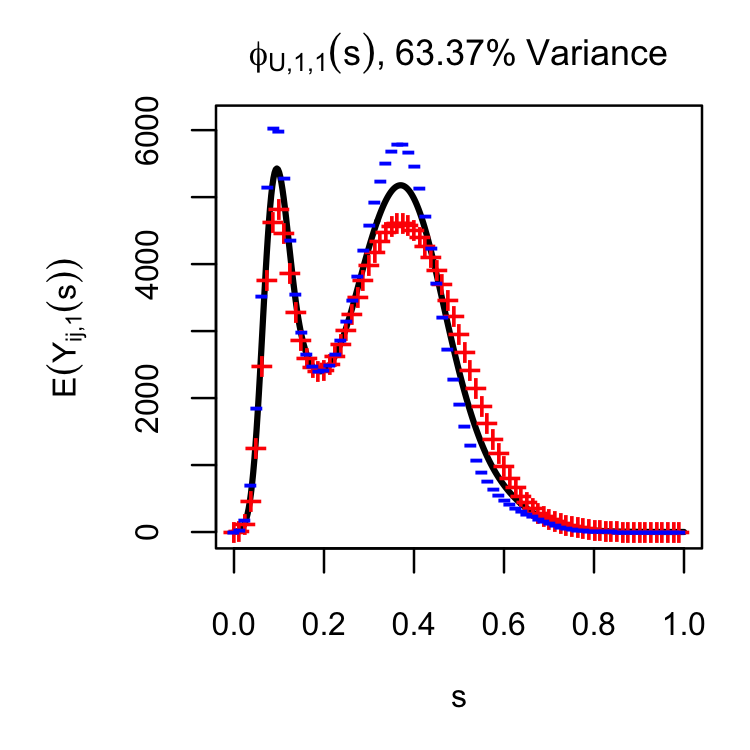}
    \end{minipage}
    \hfill
    \begin{minipage}[t]{0.49\textwidth}
        \centering
        \includegraphics[width=\linewidth]{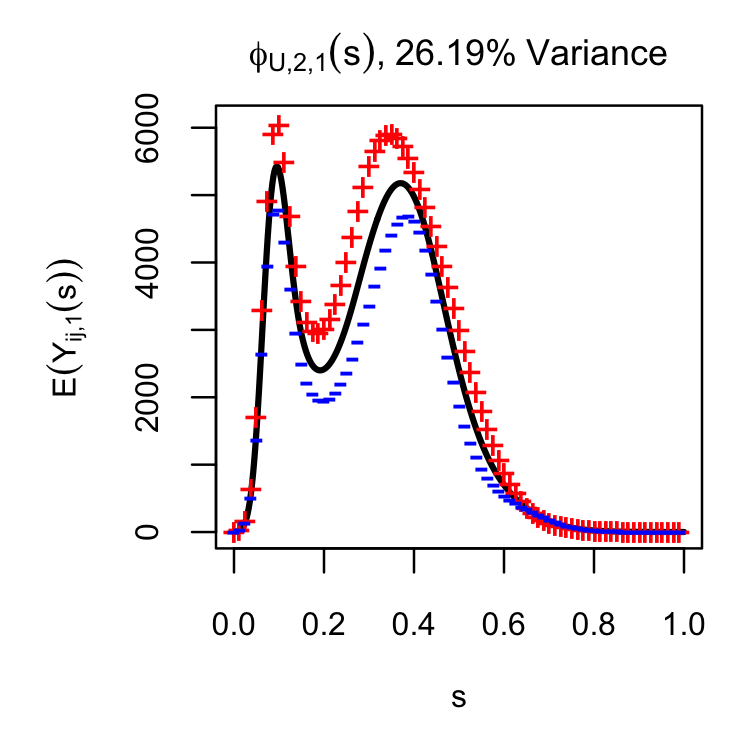}
    \end{minipage}
    \hfill
    \begin{minipage}[t]{0.49\textwidth}
        \centering
        \includegraphics[width=\linewidth]{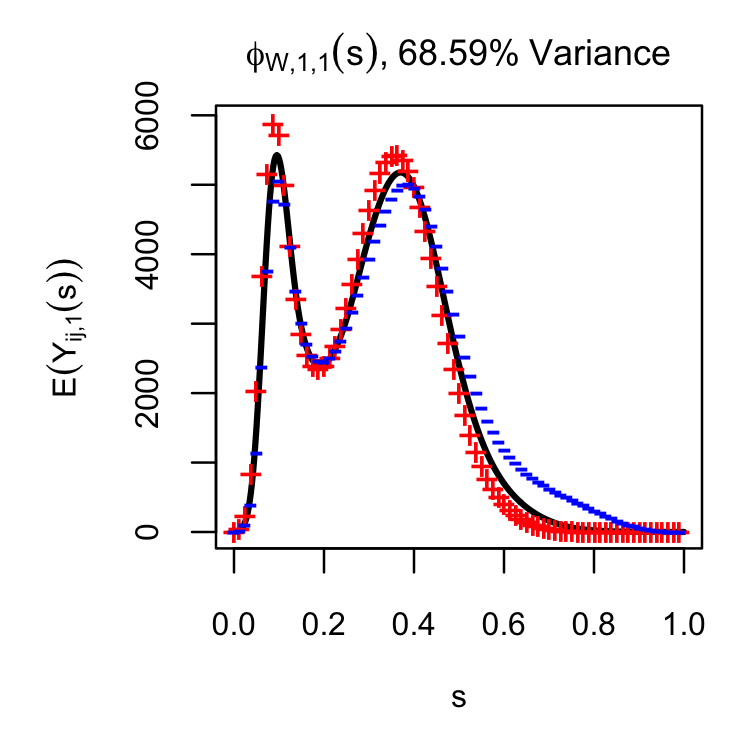}
    \end{minipage}
    \hfill
    \begin{minipage}[t]{0.49\textwidth}
        \centering
        \includegraphics[width=\linewidth]{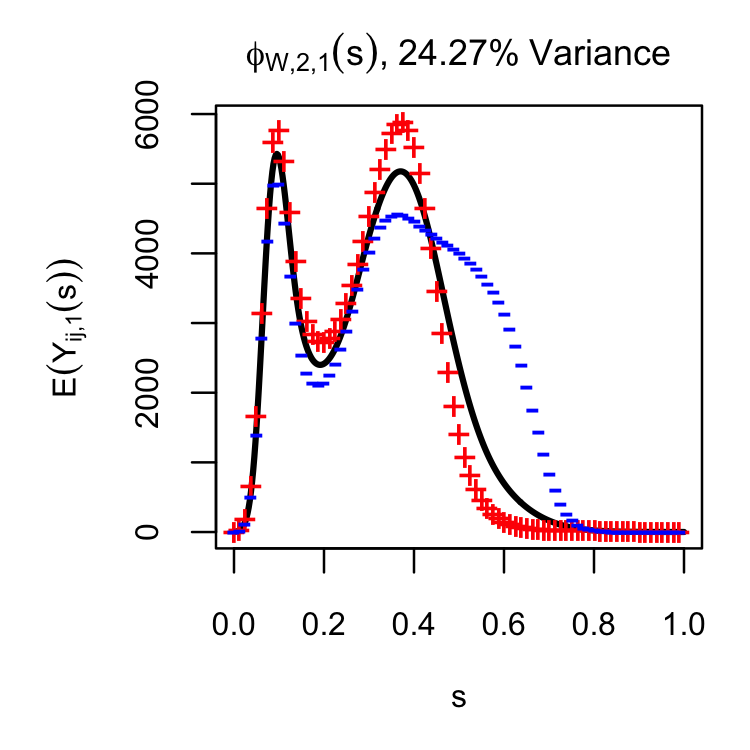}
    \end{minipage}
    \caption{Response-scale representations of the first two subject-level and visit-level functional principal components for $H_1$. The top row displays the first two subject-level components, and the bottom row displays the first two visit-level components. The black curve represents the reference expected Betti curve, while the red \(+\) and blue \(-\) symbols represent positive and negative deviations along each functional principal component direction.}
    \label{fig:fpc_h1}
\end{figure}
For \(H_1\), the first two subject- and visit-level components primarily reflect variation in the relative prominence and shape of the two major peak regions.






\end{document}